\documentclass[twocolumn,showpacs,aps,amsmath,amssymb,superscriptaddress,prl]{revtex4-1}   %showpacs,
\usepackage{amsfonts}
\usepackage{bm}
\usepackage{dcolumn}
\usepackage{setspace}
\usepackage{graphicx}
\usepackage{color}
\usepackage{multirow}
\usepackage{verbatim}
\usepackage{float}

\newcommand{\be}{\begin{equation}}
\newcommand{\ee}{\end{equation}}
\newcommand{\bea}{\begin{eqnarray}}
\newcommand{\eea}{\end{eqnarray}}
\newcommand{\bsube}{\begin{subequations}}
\newcommand{\esube}{\end{subequations}}
\newcommand{\Eq}[1]{Eq.~(\ref{#1})}
\newcommand{\Eqs}[1]{Eqs.~(\ref{#1})}

\newcommand{\ua}{\uparrow}
\newcommand{\da}{\downarrow}
\newcommand{\br}{\mathbf{r}}
\newcommand{\bR}{\mathbf{R}}

\newcommand{\ep}{\epsilon}

\def\bdu#1{\underline{\underline{\bf{#1}}}}

\begin{document}

\title{Density functional theory of electron transfer beyond the Born-Oppenheimer approximation: Case study of LiF}

\author{Chen Li}
\affiliation{Max Planck Institute of Microstructure Physics, Weinberg 2, 06120, Halle, Germany}

\author{Ryan Requist}
\affiliation{Max Planck Institute of Microstructure Physics, Weinberg 2, 06120, Halle, Germany}

\author{E. K. U. Gross}
\affiliation{Max Planck Institute of Microstructure Physics, Weinberg 2, 06120, Halle, Germany}
\affiliation{Fritz Haber Center for Molecular Dynamics, Institute of Chemistry, The Hebrew University of Jerusalem, Jerusalem 91904 Israel}

\date{\today}

\begin{abstract}

We perform model calculations for a stretched LiF molecule, demonstrating that nonadiabatic charge transfer effects can be accurately and seamlessly described within a density functional framework.  
In alkali halides like LiF, there is an abrupt change in the ground state electronic distribution due to an electron transfer at a critical bond length $R=R_c$, where a barely avoided crossing of the lowest adiabatic potential energy surfaces calls the validity of the Born-Oppenheimer approximation into doubt.
Modeling the $R$-dependent electronic structure of LiF within a two-site Hubbard model,
we find that nonadiabatic electron-nuclear coupling produces a sizable elongation of the critical $R_c$ by 0.5 Bohr.  This effect is very accurately captured by a simple and rigorously-derived correction, with an $M^{-1}$ prefactor, to the exchange-correlation potential in density functional theory; $M=$ reduced nuclear mass.  Since this nonadiabatic term depends on gradients of the nuclear wavefunction and conditional electronic density, $\nabla_R \chi(R)$ and $\nabla_R n(\mathbf{r},R)$, it couples the Kohn-Sham equations at neighboring $R$ points. Motivated by an observed localization of nonadiabatic effects in nuclear configuration space, we propose a \textit{local conditional density approximation} -- an approximation that reduces the search for nonadiabatic density functionals to the search for a single function $y(n)$.

\end{abstract}

\maketitle

\section{I. Introduction}

The many-body electron-nuclear Schr\"odinger equation is the fundamental equation of computational chemistry, but its complexity makes it difficult to find approximate solutions with ``chemical accuracy'' (1 kcal/mol $\approx$ 40~meV).  Invoking the Born-Oppenheimer (BO) approximation \cite{born1927,born1954} and working with adiabatic potential energy surfaces (PES) provides a significant simplification by effectively separating the electronic and nuclear variables.  The electronic Schr\"odinger equation with clamped nuclei can be solved by ab initio quantum chemistry methods at each point in nuclear configuration space to yield the ground state PES. The nuclear motion is characterized by the quantized vibrations and rotations on that surface.

This adiabatic treatment usually works well because the nuclear masses are significantly larger than the electron mass, rendering the nonadiabatic electron-nuclear coupling negligibly small.  However, it breaks down in several interesting cases, e.g.~when the adiabatic PES approach each other too closely, as occurs at conical intersections \cite{domcke2011}.  Non\-adiabatic effects can significantly influence chemical reactions, particularly those involving photoexcited states, proton or electron transfer, spin-orbit coupling and small energy gaps at the transition state.  Some well-known examples are alkali hydrogen halide exchange reactions (e.g.~Li$+$HF$\rightarrow$LiF$+$H) \cite{chen1980,aguado1995,ventura1996,fan2013}, collisional electron transfer reactions (e.g.~Na$+$I$\rightarrow$Na$^+$+I$^-$) \cite{moutinho1971,faist1976,kleyn1982} and reactions involving hydrogen (e.g.~F$+$H$_2$$\rightarrow$HF$+$H) \cite{tully1974,alexander2000,che2007,lique2011}.  The potential impact of nonadiabatic effects on proton transfer in water \cite{tuckerman1995,cheng1995,lobaugh1996,tuckerman1997,fang1997,decornez1999,Marx102174,cao2010,hassanali2013,rossi2016} remains largely unexplored.  A realistic description of such problems requires methods that go beyond the BO approximation.

By striking a  balance between accuracy and computational complexity, density functional theory (DFT) has become the most popular electronic structure method and perhaps the only method capable of treating large systems with quantum effects.  Therefore, it would be ideal to incorporate nonadiabatic effects into DFT.  One approach to incorporating nonadiabatic and quantum nuclear effects is to define a multicomponent DFT with both the electronic density $n(\mathbf{r})$ and $N_n$-body nuclear density $\Gamma(\bdu{R})$ as basic functional variables \cite{kreibich2001}.  As the electronic density $n(\mathbf{r})=N_e\int d\bdu{R} d\mathbf{r}_2 ...d\mathbf{r}_{N_e} |\Psi(\mathbf{r},\mathbf{r}_2,\ldots,\mathbf{r}_{N_e},\bdu{R})|^2$ is expressed in the body-fixed molecular frame and averaged over the nuclear variables $\bdu{R}\equiv(\mathbf{R}_1, \mathbf{R}_2, \cdots, \mathbf{R}_{N_n})$, it differs from the electronic density in DFT, which is a \textit{conditional} electronic density with parametric $\bdu{R}$-dependence.  Functional approximations have been introduced and tested for the hydrogen molecule \cite{kreibich2001,kreibich2008} and electron-proton correlation \cite{chakraborty2008,yang2017}, though they have not been applied to charge transfer systems.

An alternative nonadiabatic density functional theory, which works with a conditional electronic density, namely the density $n(\mathbf{r},\bdu{R}) = \langle \Phi_{\bdu{R}} | \hat{\psi}^{\dag}(\mathbf{r}) \hat{\psi}(\mathbf{r}) | \Phi_{\bdu{R}} \rangle$ calculated with the conditional electronic wavefunction $| \Phi_{\bdu{R}} \rangle$ defined in the exact factorization scheme \cite{hunter1975,gidopoulos2014,abedi2010}, has recently been proposed \cite{requist2016b}.  This theory is not a multicomponent DFT because it retains the full nuclear wavefunction $\chi(\bdu{R})$, including its gauge freedom \cite{gidopoulos2014,abedi2010}. The ground state density can be obtained by minimizing a variational energy functional.  The exact functional is not known explicitly; however, as in the Kohn-Sham (KS) scheme \cite{Koh65A1133}, one can decompose it into several components, leaving an unknown nonadiabatic Hartree-exchange-correlation (nhxc) functional $E_{nhxc}=E_{nhxc}[n,\mathbf{j}_p, \mathcal T]$ to be approximated in practice. The functional depends on two additional basic variables -- the paramagnetic current density $\mathbf{j}_p$ and the quantum geometric tensor $\mathcal{T}$ defined in Sec.~II. The functional dependence on $\mathcal T$ introduces a new complexity and properly accounting for it becomes a critical issue.

To explore the $\mathcal T$-dependence of $E_{nhxc}$, we start with a simple class of systems that often show nonadiabatic effects, namely those that experience rapid electronic density changes as the nuclear configuration is varied, implying strong electron-nuclear coupling. This reminds us of charge transfer reactions, one of the most important processes in chemistry and chemical biology.
Understanding how charge transfer takes place is a critical step towards unraveling the mechanisms of many types of reactions.
Charge transfer processes can be observed in simple diatomic molecules such as stretched LiF and NaCl \cite{Li15053001, Ruzsinszky06194112}, as well as NaI, as studied in Zewail's pioneering time-resolved vibrational spectroscopy experiments \cite{Zewail005660}.

In this paper, we use LiF as a representative charge transfer system to explore density functional approximations within the exact factorization scheme.  Instead of treating the electrons ab initio, we approximate the bond length-dependent electronic structure of LiF with an asymmetric Hubbard model, which makes the resulting equations simple enough to solve exactly.  Comparing the exact and BO solutions, we find that the major nonadiabatic effect is an elongation of the critical bond length $R_c$ at which charge transfer occurs in the conditional electronic wave function $\Phi_{\bdu{R}}$ of the molecular ground state.  We show that this effect can be accurately described by an approximation of the form $v_{nhxc}(\mathbf{r},\bdu{R}) = v_{hxc}^{\rm BO}(\mathbf{r},\bdu{R}) + v_{\rm geo}(\mathbf{r},\bdu{R})$, where $v_{hxc}^{\rm BO}(\mathbf{r},\bdu{R})$ is an hxc potential from standard DFT with parametric dependence on $\bdu{R}$ and $v_{\rm geo}(\mathbf{r},\bdu{R})$ is a geo\-metric correction that can be rigorously derived in this case from an exact nonadiabatic density functional.

The original Shin-Metiu model \cite{Shin959285}, which has been studied in the context of the exact factorization scheme \cite{abedi2013,agostini2015}, also contains charge transfer processes.  However, since that model contains only one electron, it would not allow us to study the coexistence of electron-electron correlations and nonadiabatic effects.

A different way of using DFT in conjunction with the exact factorization scheme has recently been developed in the context of a coupled-trajectory mixed quantum-classical study of quantum decoherence effects in the photochemical ring opening of oxirane \cite{min2017}.  DFT and linear response time dependent DFT were used on-the-fly to calculate the adiabatic PES and nonadiabatic coupling vectors during the self-consistent propagation of an ensemble of classical nuclear trajectories and dynamical equations for Born-Huang-like expansion coefficients describing the electronic state.  Because it employs standard DFT, which is independent of the nonadiabatic transitions that occur in the evolving state, this approach differs from exact factorization-based DFT, where the functionals themselves depend on the nonadiabaticity of the state.  Exact factorization-based DFT therefore circumvents the Born-Huang expansion and nonadiabatic coupling vectors.

The rest of the paper is structured as follows.  In section II, we briefly review the exact factorization scheme in the static case and the density functional formulation based upon it.  In section III, we apply the theory to charge transfer in the LiF molecule.  Section~IIIA motivates the use of an asymmetric two-site Hubbard model to describe the electronic structure of LiF during stretching.  The model is solved by numerical exact diagonalization in section~IIIB to provide a benchmark for subsequent density functional approximations.  The exact energy functionals within the Born-Oppenheimer approximation and within the exact factorization scheme are derived in sections~IIIC and IIID, respectively.  In section IIID, we further find that we can quantitatively capture the dominant nonadiabatic effects through a variational functional of the nuclear wave function and electronic density, without invoking the quantum geometric tensor as was proposed in our previous work \cite{requist2016b}. In section IV, we extend the formalism to general systems in continuous euclidean space with a more rigorous definition. Finally, in section V we close with some concluding remarks.

\section{II. Exact factorization scheme and density functional formulation}

Before introducing our model, let us revisit the exact factorization scheme and the density functional theory based on it.
For a nonrelativistic system of electrons and nuclei, the total Hamiltonian can be written as
\begin{equation}
    \hat H = \hat T_n + \hat H_e,
\end{equation}
where $\hat T_n$ is the nuclear kinetic energy operator and $\hat H_e=\hat T_e + \hat V_{ee}+\hat V_{en}+\hat V_{nn}$ is the Born-Oppenheimer Hamiltonian that includes electronic kinetic energy $\hat T_e$, electron-electron interaction $\hat V_{ee}$, electron-nuclear interaction $\hat V_{en}$ and nuclear-nuclear interaction $\hat V_{nn}$. The ground state of the system can be obtained through the minimization of $\langle \Psi|\hat H|\Psi \rangle$ over all possible combined electron-nuclear wave functions $\Psi(\bdu{r}, \bdu{R})$. Here we use $\bdu{r}=(\br_1, \br_2, \cdots, \br_{N_e})$ and $\bdu{R}\equiv(\bR_1, \bR_2, \cdots, \bR_{N_n})$ to denote electronic and nuclear coordinates, respectively.  The wave function can be factorized into the form $\Psi(\bdu{r}, \bdu{R})=\chi(\bdu{R})\Phi_{\bdu{R}}(\bdu{r})$ \cite{Hunter75237}, where $\chi(\bdu{R})$ is the marginal nuclear wave function and $\Phi_{\bdu{R}}(\bdu{r})$ is a conditional electronic wave function which depends parametrically on the nuclear coordinates and satisfies the partial normalization condition,
\begin{equation}
    \int |\Phi_{\bdu{R}}(\bdu{r})|^2 d\bdu{r} = 1, \quad\forall \bdu{R}.
\end{equation}
Variational determination of the ground state $\Psi(\bdu{r},\bdu{R})$ translates into the following pair of coupled equations for $\chi(\bdu{R})$ and $\Phi_{\bdu{R}}(\bdu{r})$ \cite{gidopoulos2014,abedi2010}:
\begin{align}
    \Big[\sum_{\mu=1}^{N_n} \frac{[-i\hbar \nabla_\mu+\mathbf{A}_\mu(\bdu{R})]^2}{2M_\mu}+\mathcal E(\bdu{R})\Big]\chi(\bdu{R}) &= E \chi(\bdu{R}), \label{eq-chi} \\
    \Big[\hat H_e(\bdu{R}) + \hat U_{en}^{\rm coup} [\Phi_{\bdu{R}}, \chi]\Big]\Phi_{\bdu{R}}(\bdu{r}) &= \mathcal E(\bdu{R})\Phi_{\bdu{R}}(\bdu{r}). \label{eq-phi}
\end{align}
Here $\mu$ indexes the nuclei and $M_\mu$ are the nuclear masses.  The electron-nuclear coupling gives rise to an induced vector potential
\begin{equation}
    {\mathbf{A}_{\mu}} =\langle \Phi_{\bdu{R}}| -i\hbar \nabla_\mu| \Phi_{\bdu{R}}\rangle.
\end{equation}
$\mathcal E(\bdu{R})$ is a scalar potential, defined by taking the $\bdu{r}$-space inner product of \Eq{eq-phi} with $\Phi^*_{\bdu{R}}(\bdu{r})$. Here $\hat U_{en}^{\rm coup}$ is the electron-nuclear coupling operator, given by
\begin{align}
    &\quad \hat U_{en}^{\rm coup}[\Phi_{\bdu{R}}, \chi] \nonumber \\
    &= \sum_{\mu=1}^{N_n} \frac{1}{M_\mu} \Big[ \frac{[-i\hbar \nabla_\mu - \mathbf{A}_\mu(\bdu{R})]^2}{2} \nonumber \\
    &\quad + \Big(\frac{-i\hbar \nabla_\mu \chi}{\chi}+{\mathbf{A}_\mu}(\bdu{R}) \Big)\Big(-i\hbar \nabla_\mu - \mathbf{A}_\mu(\bdu{R})\Big)\Big].
\end{align}
Solving the coupled equations \eqref{eq-chi} and \eqref{eq-phi} is completely equivalent to solving the Schr{\"o}dinger equation for the full wave function $\Psi$ and does not reduce the computational complexity.  However, it allows for further reformulation of the problem, e.g.~solving the electronic part of the problem using density functional theory.

A density functionalization of the exact factorization scheme has been proposed in Ref.~\cite{requist2016b}.  In this theory, the total energy is written as a functional of $n(\mathbf{r}, \bdu{R})$, $\mathbf{j}_p(\mathbf{r}, \bdu{R})$, $\chi(\bdu{R})$, $\mathbf{A}_\mu(\bdu{R})$ and $\mathcal T_{\mu\nu}(\bdu{R})$ as
\begin{align}
    E[n, \mathbf{j}_p, \mathcal T, \chi, \mathbf{A}] &= T_{n, \rm{marg}}[\chi, \mathbf{A}] + \int \mathcal E_{\rm geo}(\bdu{R})|\chi|^2 d\bdu{R} \nonumber \\
    &\quad + \iint  V_{en}(\mathbf{r}, \bdu{R})n(\mathbf{r}, \bdu{R}) d^3r d\bdu{R} \nonumber \\
    &\quad + \int \Big(V_{nn}(\bdu{R}) + F[n,\mathbf{j}_p, \mathcal T]\Big)|\chi|^2 d\bdu{R}. \label{orig-func}
\end{align}
Here the conditional electronic density, the paramagnetic current density and the quantum geometric tensor are defined as
\begin{equation}
    n(\mathbf{r}, \bdu{R}) = \langle \Phi_{\bdu{R}}| \hat \psi^\dagger(\mathbf{r})\hat \psi(\mathbf{r})|\Phi_{\bdu{R}}\rangle,
\end{equation}
\begin{equation}
    {\mathbf{j}_p(\mathbf{r}, \bdu{R})} = \frac{\hbar}{2im_e}\langle \Phi_{\bdu{R}}| \hat \psi^\dagger(\mathbf{r})\nabla\hat \psi(\mathbf{r})-\nabla\hat \psi^\dagger(\mathbf{r})\hat \psi(\mathbf{r})|\Phi_{\bdu{R}}\rangle,
\end{equation}
with $m_e$ being the electron mass, and
\begin{equation}
    \mathcal T_{\mu\nu} = \langle \partial_\mu \Phi_{\bdu{R}}|(1-| \Phi_{\bdu{R}}\rangle \langle \Phi_{\bdu{R}}|)|\partial_\nu \Phi_{\bdu{R}} \rangle.
\end{equation}
In \Eq{orig-func}, $T_{n, \rm{marg}}$ is the marginal nuclear kinetic energy,
\begin{align}
    T_{n, \rm{marg}} = \int \chi^*(\bdu{R})\sum_{\mu=1}^{N_n} \frac{[-i\hbar \nabla_\mu+\mathbf{A}_\mu(\bdu{R})]^2}{2M_\mu}\chi(\bdu{R}) d\bdu{R}.
\end{align}
$\mathcal E_{\rm geo}(\bdu{R})$ is a geometric contribution to the energy \cite{Requist16042108},
\begin{equation}
    \mathcal E_{\rm geo}(\bdu{R}) = \frac{\hbar^2}{2} I^{\mu\nu} \mathcal T_{\mu\nu},
    \label{Egeo}
\end{equation}
with $I^{\mu\nu}=\delta^{\mu\nu}/M_{\mu}$ being an inverse inertia tensor.  In \Eq{orig-func}, $F$ is an electronic functional implicitly defined through a constrained search as
\begin{equation}
    F[n,\mathbf{j}_p, \mathcal T] = \min_{\Psi \rightarrow (n,\mathbf{j}_p, \mathcal T)} \langle \Phi_{\bdu{R}}|\hat T_e + \hat V_{ee} |\Phi_{\bdu{R}}\rangle {,}
\end{equation}
which is universal in the sense that it does not depend on $\chi$ or $\hat{V}_{en}$.
The minimization of $E[n, \mathbf{j}_p, \mathcal T, \chi, \mathbf{A}]$ can be reduced to solving (i) the Schr{\"o}diner equation for $\chi(\bdu{R})$, (ii) conditional Kohn-Sham equations for $n(\mathbf{r}, \bdu{R})$ and $\mathbf{j}_p(\mathbf{r}, \bdu{R})$, and (iii) an Euler-Lagrange equation for $\mathcal T_{\mu\nu}(\bdu{R})$. The validity of this framework has been demonstrated for the $E\otimes e$ Jahn-Teller model. However, due to the one-electron nature of that model, the electronic functional $F$ reduces to the noninteracting electronic kinetic energy $T_{e,s}$, and thus $E_{nhxc}[n,\mathbf{j}_p, \mathcal T]=F[n,\mathbf{j}_p, \mathcal T]-T_{e,s}[n,\mathbf{j}_p]$ vanishes identically.  Therefore, one cannot use the $E\otimes e$ Jahn-Teller model to study the functional form of $E_{nhxc}$.  For many-electron systems, the form of $E_{nhxc}$, particularly its dependence on $\mathcal T$ remains unknown.

\section{III. Application to L\MakeLowercase{i}F}

To explore how the quantum geometric tensor can be accounted for in many-electron systems, we start by considering simple diatomic molecules that show nontrivial nonadiabatic effects.  A candidate system with relatively light nuclei is the LiF molecule, where charge transfer takes place when the bond is stretched beyond a critical value.  To simplify the full problem in three-dimensional space, we assume that both nuclei are constrained to lie along a laboratory-fixed axis.  Hence, we neglect the rotational degrees of freedom and rovibronic coupling, and after separating off the nuclear center of mass motion, only a single nuclear variable remains -- the bond length $R$.  Since the nuclear configuration space is one-dimensional, a gauge can be chosen that eliminates the induced vector potential $\mathbf{A}_\mu(\bdu{R})$.  Moreover, the paramagnetic current density $\mathbf{j}_p(\mathbf{r})$ must also vanish for the ground state.  This enables us to focus on the functional dependencies on $\chi$, $n$ and $\mathcal T$.

With these assumptions, the electron-nuclear Hamiltonian reduces to
\begin{equation}
    \hat{H}(\bdu{r}, R) = -\frac{\hbar^2}{2M} \frac{d^2}{dR^2} + \hat H_e(\bdu{r}, R) {,}
\end{equation}
where $M=\frac{M_1 M_2}{M_1+M_2}$ is the reduced nuclear mass and $\hat H_e$ depends only on the bond length $R=|R_1-R_2|$; we assume $R_1\leq R_2$, and let $R_1$ refer to the position of the Li atom and $R_2$ to that of the F atom.  The full electron-nuclear wave function is the solution of the Schr{\"o}dinger equation
\begin{equation}
    \hat H(\bdu{r}, R) \Psi(\bdu{r},R) = E \Psi(\bdu{r},R). \label{Hv}
\end{equation}

We transform all units to atomic units so that $\hbar =1$.  In terms of the proton mass $m_p$, the two nuclear masses are $M_1=7m_p$ and $M_2=19m_p$ (here we treat the proton and neutron masses as identical) so that $M=5.1154m_p$.  One can further transform the reduced mass into atomic units, giving $M=9392m_e$. 
% and $\kappa = (m_e/M)^{1/4} \approx 0.10$.

\subsection{IIIA. Two-site Hubbard model for the electrons}

Although the nuclear part of the problem defined by \Eq{Hv} is manageable, the electronic Hamiltonian $\hat H_e$ is too complicated to solve exactly.  It also poses a challenge to state-of-the-art electronic density functionals \cite{Li15053001}.  To simplify the electronic Hamiltonian while keeping the essential charge transfer physics, we consider only the two valence electrons involved in the chemical bond, i.e.~the $2s$ electron of Li and the unpaired $2p$ electron of F. This reduces the problem to an asymmetric two-site Hubbard model with 2 electrons.

The Hamiltonian of the two-site Hubbard model is
\begin{equation}
    \hat H_e =  -t \sum_{\sigma}(\hat c^\dagger_{1\sigma} \hat c_{2\sigma} + \hat c^\dagger_{2\sigma}\hat c_{1\sigma})  + \sum_i U_i \hat n_{i\ua} \hat n_{i\da} +\sum_{i} \ep_{i} \hat n_{i},
\end{equation}
where $\hat c^\dagger_{i\sigma}$, $\hat c_{i\sigma}$ and $\hat n_{i\sigma}$ are creation, annihilation and electron number operators for spin $\sigma$ on site $i$; $\hat n_i = \sum_\sigma \hat n_{i\sigma}$.  The three terms on the right hand side (rhs) represent electron hopping, on-site Hubbard interactions and on-site potential energy (assumed to be spin-independent). The electron-nuclear attraction and internuclear repulsion energies have been effectively absorbed into the first and last terms.  We have assumed that the Hubbard interactions $U_i$ and the on-site energies $\ep_i$ are site-dependent. For simplicity, we restrict to three singlet states, namely, $\varphi_1 = |1_\ua 1_\da\rangle$, $\varphi_2 = \frac{1}{\sqrt 2}(|1_\ua 2_\da\rangle-|1_\da 2_\ua\rangle$), and $\varphi_3 = |2_\ua 2_\da\rangle$.  In the representation of these basis states, the model Hamiltonian becomes
\begin{equation}
\bm H_e=
  \begin{bmatrix}
    2\ep_1+U_1 & -\sqrt 2t & 0  \\
    -\sqrt 2t & \ep_1+\ep_2 & -\sqrt 2t \\
    0 & -\sqrt 2t & 2\ep_2 + U_2
  \end{bmatrix}.
\end{equation}
Denoting $\bm E_0 = (\ep_1+\ep_2)\bm I \equiv \ep_0 \bm I$ with $\bm I$ being the identity matrix and subtracting $\bm E_0$ from $\bm H_e$, gives the following Hamiltonian:
\begin{equation}
\bm {\bar H}_e= \bm H_e -\bm E_0 =
  \begin{bmatrix}
    U_1 + \Delta \ep & -\sqrt 2t & 0  \\
    -\sqrt 2t & 0 & -\sqrt 2t \\
    0 & -\sqrt 2t & U_2 - \Delta \ep
  \end{bmatrix}, \label{bar-He}
\end{equation}
where $\Delta \ep = \ep_1 - \ep_2$.

In applications of the two-site Hubbard model targeting a particular molecular geometry, the parameters $t$, $U_i$ and $\ep_i$ can be taken to be numbers. However, since our aim is to model the coupling between the electronic state and the bond length, we have to consider these parameters as $R$-dependent functions. In the dissociation limit $R\rightarrow \infty$, the parameters approach the following limiting values: $t\rightarrow 0$; $U_i=I_i-A_i$, where $I_i$ is the ionization potential (IP) and $A_i$ is the electron affinity (EA) of atom $i$; and $\ep_i=-I_i$.  By looking up the experimental IP and EA of Li and F, we can calculate these parameters as listed in Table \ref{IP-EA}.

\setlength{\tabcolsep}{11pt}
\setlength{\extrarowheight}{2pt}

\begin{table}[H]
\caption{IP, EA and Hubbard model parameters of Li and F atoms when $R\rightarrow\infty$. All values are in eV. \label{IP-EA}}
\centering
\begin{tabular}{ccccc}
\hline\hline
 & IP & EA & $U_i(R\rightarrow \infty)$ & ${\ep_i}(R\rightarrow \infty)$\tabularnewline
\hline\hline
Li & 5.39 & 0.62 & 4.77 & -5.39\tabularnewline
F & 17.42 & 3.40 & 14.02 & -17.42\tabularnewline
\hline\hline
\end{tabular}
\end{table}

We choose the energy reference by setting the energy of the total system to be zero in the dissociation limit $R\rightarrow \infty$; one can verify that this energy is given by $\ep_0^\infty = \ep_0(R=\infty)$.  Therefore, we introduce the Hamiltonian
\begin{equation}
\bm{\tilde H}_e = \bm H_e - \ep_0^\infty \bm I = \bm {\bar H}_e + (\ep_0-\ep_0^\infty)\bm I {.} \label{HTilde}
\end{equation}

To choose the $R$-dependence of $t$, $\Delta \ep$ and $\tilde{\ep}_0 \equiv \ep_0-\ep_0^\infty$, we start by analyzing their large-$R$ asymptotic behavior (here we ignore the $R$-dependence of $U_i$ for simplicity). First of all, $t$ is a hopping integral between atomic orbitals $\phi_i$ on different sites. For large $R$, the two orbitals can be considered as proportional to two exponentially decaying functions centered at the two nuclei and separated by $R$. Thus $t$ is expected to decay exponentially as a function of $R$.  Hence we model this term by
\begin{equation}
    t = t_0 e^{-\beta R}, \label{hopping}
\end{equation}
where $t_0$ and $\beta$ are constant parameters to be fixed below.

When $R\rightarrow \infty$, $\Delta \ep$ is given by the difference between the IP of the two sites, i.e., $\Delta \ep = \Delta I=I_2-I_1$. For finite $R$, the presence of the other atom leads to a correction to the IP of each site, and hence a correction to $\Delta \ep$. By performing a multipole expansion, one can derive that the leading order terms in $1/R$ are given by \cite{Supp}
\begin{equation}
    \Delta \ep = \Delta I + \frac{\gamma}{R^3}, \label{Delta-ep-1}
\end{equation}
where $\gamma$ is a parameter related to the quadrupole moment integral. \Eq{Delta-ep-1} has an unphysical singularity at $R=0$. To remove this artifact, we introduce a parameter $R_0$ in the denominator,
\begin{equation}
    \Delta \ep = \Delta I + \frac{\gamma}{R^3+R_0^3}, \label{Delta-ep-2}
\end{equation}
so that $\Delta \ep$ is finite at $R=0$.

Finally, $\tilde{\ep}_0$ determines the overall shape of the PES. Since dissociation energy curves of diatomic molecules can be modeled by the Morse potential, here we also write $\tilde{\ep}_0$ as a Morse potential,
\begin{equation}
    \tilde{\ep}_0 = D_e \Big[ e^{-2\alpha(R-R_e)}- 2e^{-\alpha(R-R_e)} \Big], \label{bar-ep0}
\end{equation}
where $R_e$ is the equilibrium bondlength; $D_e$ is the well depth and $\alpha$ controls the width. It is worth remarking that the choice of $D_e$ and $\alpha$ is closely connected with the binding energy and the well width predicted by the two-site Hubbard model, although not exactly the same.

To realistically model the molecular dissociation curve, we take the results of ab initio calculations \cite{Li15053001} using coupled cluster with singles, doubles and perturbative triples, CCSD(T) \cite{Cizek69,Purvis82,Pople87}, as the benchmark and fit our undetermined parameters so that the binding energy, charge transfer position and overall shape are reproduced. We have not considered excited state PES in our fitting so that the excited state PES predicted by our model has an unphysical well near the minimum $R$; this has, however, no relevance for our present ground state calculations.  It is possible to correct the deficiency and accurately model multiple PES in our model by better characterizing the model parameters in the small $R$ region. For example, refining the $R$-dependence of $\Delta \ep$ and considering $R$-dependent $U_i$ will improve the model.  Nevertheless, since we are focusing on the ground state in this paper, we content ourselves with a minimal model that is able to reproduce the ground state PES as well as the excited state PES in the avoided crossing region.

\begin{figure*}[t]
\includegraphics[width=0.75\textwidth]{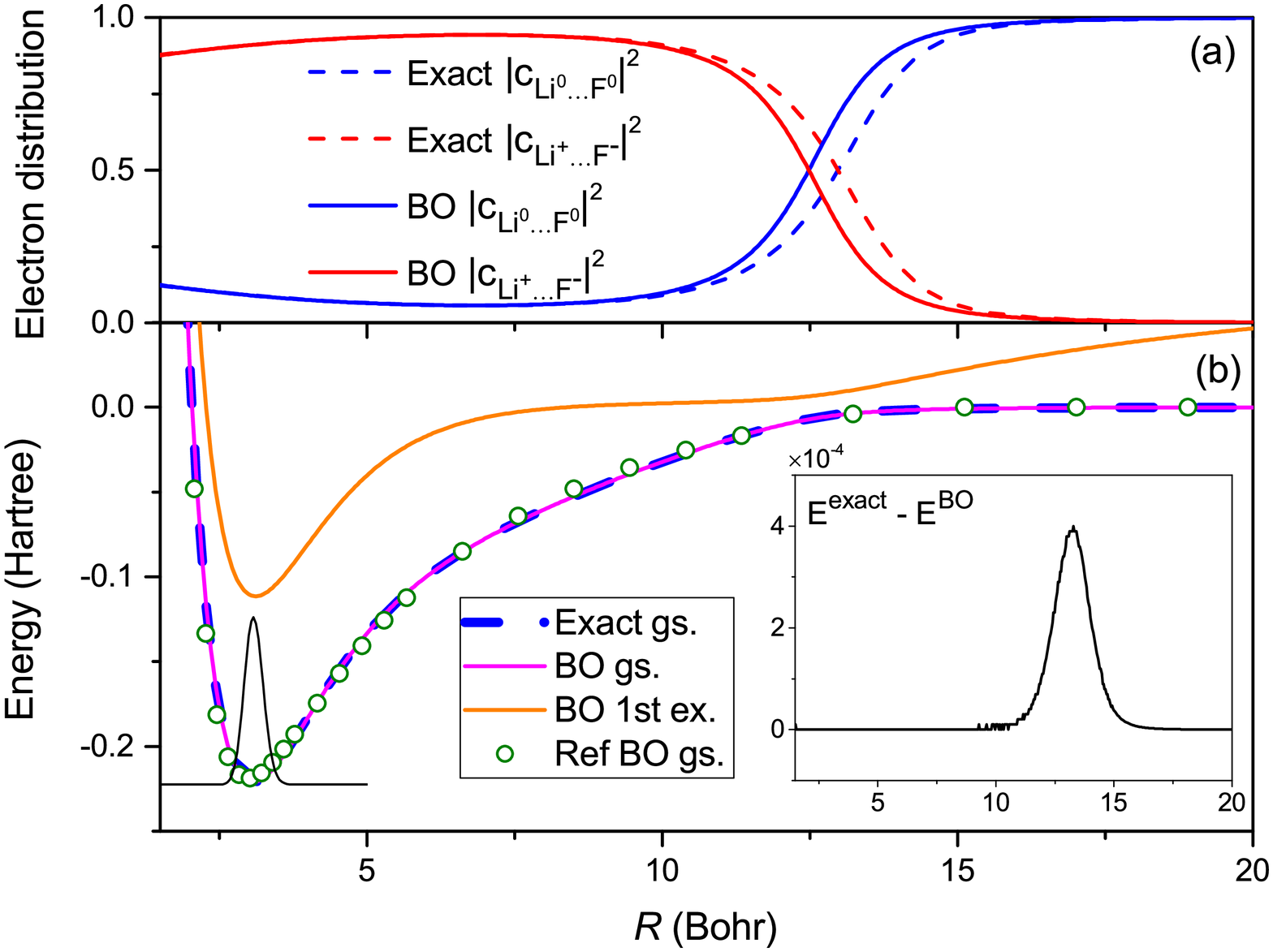}
\caption{
(a) Populations of the many-body configurations in the conditional electronic wave functions $|\Phi_R\rangle$ and $|\Phi_R^{\rm BO}\rangle$; a third higher-energy state $|c_{\rm Li^-\cdots \rm F^+}|^2$ is negligible for all $R$ and not shown. (b) Comparison between the exact and BO ground state potential energy surfaces in our model.  Also shown are reference ab initio coupled cluster data for LiF in the BO approximation and the first excited state BO potential energy surface in our model. \label{fig1}}
\end{figure*}

Our fit to the ground state PES is shown in Fig.~\ref{fig1}(b).
The parameters are $t_0=1$eV, $\beta=0.163$ Bohr$^{-1}$, $\gamma=255$ Hartree$\times$Bohr$^3$, $R_0=11.5$ Bohr, $R_e=3.1$ Bohr, $D_e=0.12$ Hartree and $\alpha=0.8152$ Bohr$^{-1}$.
As can be seen, the ground state BO-PES in the two-site Hubbard model, obtained from the smallest eigenvalue of $\bm {\tilde H}_e$ as a function of $R$, is a remarkably accurate fit to the ab initio result; the agreement is semi-quantitative.
Moreover, the charge transfer position (where the percentage of the ionic species Li$^+\cdots$F$^-$ becomes identical with that of the neutral species Li$^0\cdots$F$^0$) is around $R=12.5$ Bohr.
This validates our two-site Hubbard model as a good starting point to study nonadiabatic effects in the LiF molecule.

\subsection{IIIB. Solution of the full Schr\"odinger equation}

Having restricted the electrons to a three-dimensional Hilbert space, the full Schr\"odinger equation we want to solve is
\begin{equation}
    [-\frac{1}{2M}\frac{d^2}{dR^2} +\bm{\tilde H}_e(R)]\Psi(R) = E \Psi(R), \label{exact-Sch}
\end{equation}
where $\Psi(R) = [a_1(R), a_2(R), a_3(R)]^T$.  To solve this equation numerically, we expand $\Psi(R)$ as
\begin{equation}
    \Psi(R) = \sum_{nk} C_{nk} B_n(R)\hat e_k.
\end{equation}
Here, we adopt the B-spline functions $\{B_n(R)\}$ as real space basis functions (see Ref \cite{Hofmann01245321, Supp}) and $\hat e_k$ as three-component electronic basis vectors; the $k$th component of $\hat e_k$ is 1 and the rest are 0.  This transforms \Eq{exact-Sch} into an algebraic eigenvalue equation, by which we can solve for the ground state energy and wave function.
With the ground state electronic wave function available, we can evaluate the exact PES and perform a population analysis of the electronic states.  First, we write $\Psi(R)$ in its exact factorized form $\Psi(R) = \chi(R)\Phi(R)$, where $\chi(R)=\sqrt{a_1(R)^2+a_2(R)^2+a_3(R)^2}$ is the nuclear wave function for the vibrational degree of freedom and $\Phi(R) = [c_1(R), c_2(R), c_3(R)]^T$ is the conditional electronic wave function with $c_i(R)=a_i(R)/\chi(R)$.

In Fig.~\ref{fig1}, the exact PES and the exact populations of neutral and ionic configurations, $|c_2(R)|^2$ and $|c_3(R)|^2$, are compared with the corresponding BO results.  The exact ground state surface almost coincides with the BO one, with the energy difference on the magnitude of $10^{-4}$ Hartree as shown in the inset of Fig \ref{fig1}(b). However, there is a qualitative difference in the electronic populations in the range of 10 to 15 Bohr. In particular, in the exact solution, the charge transfer bond length is $R_c\approx 13.0$ Bohr, about 0.5 Bohr longer than the BO prediction.
This is a nonadiabatic effect:
as the bond is stretched, the coupling between nuclear and electronic wave functions---beyond what is already present in the BO approximation---causes the electron transfer to occur at a longer internuclear distance.  The electronic populations are in good qualitative agreement with populations inferred from ab initio calculations of electrical dipole moments in LiF \cite{bauschlicher1988,hanrath2008} and LiCl \cite{weck2004,kurosaki2012}.

\subsection{IIIC. Born-Oppenheimer-based density functional}

In this section, we numerically derive the density functional for the asymmetric two-site Hubbard model in the BO approximation.  Varying the on-site potentials $\epsilon_i$ in a Hubbard model and solving the Schr\"odinger equation allows one to define a mapping $\{\epsilon_i\} \rightarrow \{n_i\}$, which is analogous to the $v(\mathbf{r})\rightarrow n(\mathbf{r})$ mapping in standard DFT.  Here, $\{n_i\}$ are the site occupation numbers, and one can construct \textit{site occupation} functionals, e.g.~$E_{xc}[\{n_i\}]$ \cite{schoenhammer1995}.  The simplest example is the two-site Hubbard model, the many applications of which are reviewed in Ref.~\cite{Carrascal15393001}.

To construct the ground state energy functional for our two-site Hubbard model in the BO approximation, we first observe that for any  $R$, the BO electronic wave function $\Phi^{\rm BO}$ can be parameterized by two variables $\theta_1$ and $\theta_2$ as $\Phi^{\rm BO}=(\cos \theta_1 \sin \theta_2, \sin \theta_1, \cos\theta_1 \cos\theta_2)^T$. For the remainder of this section,
the parametric $R$-dependence of the variables, which should not be confused with the $\mathbf{r}$-dependence of the density $n(\mathbf{r})$ in DFT, is suppressed for brevity.  It follows that the electronic energy can be written in terms of $\theta_1$ and $\theta_2$ as
\begin{align}
    E_e[\theta_1, \theta_2] &= \langle \Phi | \tilde{\bm H}_e |\Phi\rangle \nonumber \\
    &= -\sqrt 2 t \sin 2\theta_1 (\sin \theta_2 + \cos \theta_2) \nonumber \\
    &\quad + \cos^2 \theta_1 (\tilde U_1 \sin^2 \theta_2 + \tilde U_2 \cos^2 \theta_2) + \tilde{\epsilon}_0. \label{energy}
\end{align}
Here $\tilde U_1 = U_1 + \Delta \epsilon$ and $\tilde U_2 = U_2 - \Delta \epsilon$.

Now we introduce the ``electron density'' $n$ as the population difference between site 2 and site 1, i.e.,
\begin{equation}
    n = \cos^2\theta_1 \cos^2\theta_2 - \cos^2 \theta_1 \sin^2 \theta_2 = \cos^2 \theta_1 \cos 2\theta_2, \label{density}
\end{equation}
which ranges from -1 to 1. We will neglect the subtle distinctions between site occupation functional theory and DFT and adopt $n$ in \Eq{density} as our density variable.

Following the constrained search formulation \cite{Levy796062}, we define the density functional as
\begin{equation}
    E_e^{\rm BO}[n] = \lim_{\theta_1, \theta_2 \rightarrow n} E_e[\theta_1, \theta_2]. \label{Levy}
\end{equation}
The first term (hopping term) on the rhs of \Eq{energy} is small compared to the others.
If we neglect this term, we eliminate the $\theta_2$ dependence of $E_e$ and minimizing the resulting function of $\theta_1$ over the domain $\cos^2\theta_1 \geqslant |n|$ leads to $\cos^2\theta_1=|n|$ and $|\cos2\theta_2| = 1$, i.e.~the minimum is achieved at the boundary of the $(\theta_1,\theta_2)$ domain.
Assuming that the minimizer is pinned to the boundary in \Eq{energy}, one can deduce the approximate functional
\begin{align}
    E_e^{\rm BO,approx}[n] &= -2\sqrt 2 t \sqrt{|n|(1-|n|)}+\frac{1}{2}|n| (\tilde U_1+\tilde U_2)\nonumber \\
    &\quad + \frac{1}{2}n(\tilde U_2-\tilde U_1) + \tilde{\epsilon}_0.
\end{align}
To quantify the deviation of $\theta_1$ from its boundary, we introduce a new variable \begin{equation}
    u = \sqrt{1-\frac{|n|}{\cos^2\theta_1}}\;,
\end{equation}
which ranges from 0 to $\sqrt {1-|n|}$. The pair of variables $(n,u)$ essentially contains the information of $(\theta_1, \theta_2)$ through a variable transformation. By this change of variables, we can rewrite the exact BO functional in terms of a one-dimensional minimization over $u$ as \cite{Supp}
\begin{align}
    E_e^{\rm BO}[n] &= \min_{0\leqslant u \leqslant \sqrt{1-|n|}} \Big\{-2\sqrt 2 t\sqrt{\bigg(1-\frac{|n|}{1-u^2}\bigg)\frac{|n|}{1-u^2}}\nonumber \\
    &\quad \times \sqrt{1+ u\sqrt{2-u^2}\;}+ \frac{|n|}{2(1-u^2)}(\tilde U_1+\tilde U_2) \Big\}\nonumber \\
    &\quad + \frac{1}{2}n(\tilde U_2-\tilde U_1) + \tilde{\epsilon}_0, \quad (n\neq0)\label{exact_func}
\end{align}
and
\begin{align}
    E_e^{\rm BO}[0] &= -\sqrt{4t^2+\frac{1}{16}(\tilde U_1+\tilde U_2)^2}+ \frac{1}{4}(\tilde U_1+\tilde U_2)+ \tilde{\epsilon}_0. \label{energy-1}
\end{align}
$E_e^{\rm BO,approx}[n]$ is an excellent approximation to $E_e^{\rm BO}[n]$; the major deviation is near $n=0$, where the maximum error for most $R$
is on the order of $10^{-3}$ Hartree \cite{Supp}.

\begin{figure}[t]
\includegraphics[width=\columnwidth]{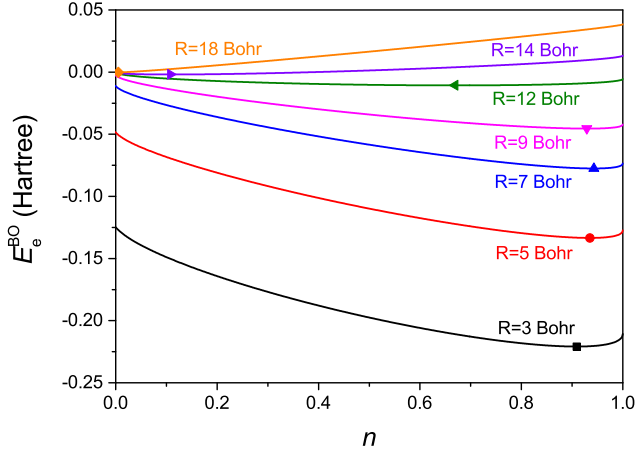}
\caption{
Ground state density functional within BO approximation for different $R$. Here we focus on the range of $n\in[0,1]$ and the global minimum of each curve (whose coordinates represent the ground state density and energy for each $R$) has been marked in the plot.  \label{fig2}}
\end{figure}

Figure \ref{fig2} shows $E_e^{\rm BO}$ on the domain $[0,1]$ for a series of $R$.  Near the equilibrium bond length ($R\approx 3$~Bohr), the minimum occurs around $n=0.9$. As $R$ increases, the energy curve rises up and deforms into a shallower shape. Moreover, the minimum at first slides towards $n=1$ and but then changes direction and begins to slide back for $R\gtrsim 7$~Bohr. When $R$ reaches a critical value ($R\approx 12$~Bohr), the minimum shifts abruptly to a value very close to $n=0$. This is consistent with our observation of a charge transfer around that distance.
Plots for the whole domain $[-1,1]$ and a comparison between $E_e^{\rm BO}$ and $E_e^{\rm BO,approx}$ can be found in the Supplemental Material.

The charge transfer occurs due to the competition between the on-site potential difference $\Delta \epsilon$ and the Hubbard interactions; when the molecule is stretched beyond $R\gtrsim 12$~Bohr, interactions win and the system snaps into the neutral configuration with a single electron occupying each site.  In the symmetric Hubbard model, the ratio of $U/t$ determines the strength of correlations in the system. This measure, however, cannot be directly applied to our asymmetric model. Because $t$ is small, the $3\times3$ reduced Hamiltonian in \Eq{bar-He} can be approximately treated as a block diagonal matrix, with the two blocks being $U_1+\Delta \ep$ and $[0,-\sqrt 2 t; -\sqrt 2 t, U_2-\Delta \ep]$. The second block effectively reduces to a Hubbard Hamiltonian involving states $|\varphi_2\rangle$ and $|\varphi_3\rangle$. Therefore the effective ratio $q(R)=\frac{U_2-\Delta \ep(R)}{\sqrt 2 t(R)}$, which can take negative values for $R<R_c$, predicts the amount of correlation in our system: the system is weakly correlated for $R<R_c$ but becomes increasingly strongly correlated when $R>R_c$. This is discussed in further detail in the Supplemental Material \cite{Supp}, where we analyze the natural occupation numbers as functions of $R$.

\subsection{IIID. Exact factorization-based density functional}

The functional $E_e^{\rm BO}[n]$ defined in the previous section does not contain nonadiabatic effects.  Since the exact factorization scheme \cite{gidopoulos2014, abedi2010} lends itself to the definition of a \textit{conditional electronic density} $n(\mathbf{r},\bdu{R})$ that includes all nonadiabatic effects, it provides a rigorous foundation for a beyond-BO density functional theory \cite{requist2016b} in which the variational energy minimization yields the density $n(\mathbf{r},\bdu{R})$ instead of the BO density $n^{\rm BO}(\mathbf{r},\bdu{R})$ -- the latter conventionally denoted $n(\mathbf{r})$.  The beyond-BO electronic energy is generally also a functional of the paramagnetic current $\mathbf{j}_p(\mathbf{r},\bdu{R})$ and quantum geometric tensor $\mathcal{T}_{\mu\nu}(\bdu{R})$.

In this section, we explore how to express the electronic energy functional for our model in terms of the conditional electronic density $n(R)$ and the quantum geometric scalar $g(R)$ (the tensor $\mathcal T_{\mu\nu}$ reduces to a scalar $g(R)$ since the nuclear configuration space is one-dimensional). Then, motivated by the observation that $g(R)$ is approximately redundant with $n(R)$, we express the electronic energy as a functional of $n(R)$ alone.

In our model, $\Phi(R)$ is parameterized by $\theta_1$ and $\theta_2$, which through a coordinate transformation can be written as a function of $n(R)$ and the auxiliary variable $u(R)$. Therefore, the total energy is a functional of $n(R)$, $u(R)$ and $\chi(R)$,
\begin{align}
    &\quad E[n(R), u(R), \chi(R)] \nonumber \\
    &= -\frac{1}{2M}\int \chi^*(R)\nabla^2\chi(R)dR + E_e[n(R), u(R), \chi(R)],
\end{align}
where
\begin{equation}
    E_e[n(R), u(R), \chi(R)] = \int |\chi(R)|^2 \Big[E_{e}[n,u]+\frac{g(R)}{2M}\Big]dR. \label{Exfunc-1}
\end{equation}
Here $E_{e}[n,u]$ is the quantity in braces on the rhs of \Eq{exact_func}, which is $E_e[\theta_1,\theta_2]$ in \Eq{energy} expressed in terms of $n(R)$ and $u(R)$, and
\begin{equation}
    g(R) = \left< \frac{d\Phi}{dR} \bigg| \frac{d\Phi}{dR} \right> = \sum_{i=1}^3 \bigg(\frac{d c_i}{dR}\bigg)^2.
\end{equation}
The function $g(R)$ can be recast into an expression of $n(R)$, $u(R)$ and their derivatives as follows \cite{Supp}:
\begin{align}
    g(R) &= C_{nn}\Big(\frac{dn}{dR}\Big)^2 + C_{uu} \Big(\frac{du}{dR}\Big)^2 + C_{nu}\frac{dn}{dR}\frac{du}{dR},
    \label{g-1}
\end{align}
where
\begin{equation}
    C_{nn} = \frac{1}{4n(1-u^2-n)},
\end{equation}
\begin{equation}
    C_{uu} = \frac{n(1+nu^2-n)}{(1-u^2)^2(1-u^2-n)(2-u^2)},
\end{equation}
and
\begin{equation}
    C_{nu} = \frac{u}{(1-u^2)(1-u^2-n)}.
\end{equation}
If we assume $n(R)$ and $g(R)$ are known functions and solve the differential equation in \Eq{g-1} for the unknown function $u(R)$ with the appropriate boundary conditions, we define a functional $u[n(R),g(R)]$, which, when substituted back into \Eq{Exfunc-1}, formally defines an electronic functional $E_{e}[n(R),g(R)]$. However, since to obtain an explicit form we would have to be able to solve \Eq{g-1} for \textit{arbitrary} $n(R)$ and $g(R)$, which is mathematically challenging, we here follow a different strategy that additionally allows us to eliminate the functional dependence on $g(R)$.

First, we observe that the $C_{nn}$ term in \Eq{g-1} is dominant for all $R$ (see the plots of the individual terms in the Supplemental Material \cite{Supp}).  Moreover, since $u(R)$ is small in most regions of interest, we can drop the $u$-dependence in $C_{nn}$ so that in this approximation $g(R)$ depends only on $n(R)$ and $dn(R)/dR$, i.e.
\begin{align}
    g(R) &\approx \frac{1}{4n(1-n)}\Big(\frac{dn}{dR}\Big)^2.
    \label{g-2}
\end{align}
Substituting \Eq{g-2} into \Eq{Exfunc-1}, and replacing $E_{e}[n,u]$ by $E_e^{\rm BO,approx}[n]$ (i.e., setting $u$ to be zero), we arrive at a functional that depends only on $n(R)$ and $\chi(R)$,
\begin{align}
    E_e[n(R), \chi(R)] &= \int dR |\chi(R)|^2 \times \nonumber \\
    &\Big[E_{e}^{\rm BO,approx}[n] +\frac{f(n)}{2M}\Big(\frac{dn}{dR}\Big)^2\Big], \label{Exfunc-2}
\end{align}
with $f(n) = \frac{1}{4n(1-n)}$.

We refer to \Eq{Exfunc-2} together with \Eq{g-2} as the local conditional density approximation (LCDA), since (i) it reduces the full electronic part of the functional that depends on the complete information of $\Phi(R)$ to a much simpler one that depends on the conditional electronic density and its $R$-space gradient; and (ii) the prefactor $f(n)$ is local in $n(R)$.

The variation of \Eq{Exfunc-2} with respect to $n(R)$ and $\chi(R)$ (subject to a normalization constraint) leads to coupled Euler-Lagrange (EL) equations, which after simplification read
\begin{equation}
    -\frac{1}{2M}\nabla^2 \chi + \Big[E_{e}^{\rm BO,approx}[n]+\frac{f(n)}{2M}\Big(\frac{dn}{dR}\Big)^2\Big]\chi = E \chi, \label{EL-chi2}
\end{equation}
\begin{align}
     \frac{d E_{e}^{\rm BO,approx}}{dn} + v_{\rm geo}\Big[n, \frac{dn}{dR}, \frac{d\big(\rm ln |\chi|^2\big)}{dR}\Big]=0, \label{EL-n2}
\end{align}
where
\begin{align}
    v_{\rm geo}(R) &= -\frac{1}{M}\Big[\frac{1}{2}f'(n)\Big(\frac{dn}{dR}\Big)^2+f(n)\Big(\frac{d^2n}{dR^2}\Big) \nonumber \\
    &\quad+\frac{d\Big(\rm ln |\chi(R)|^2\Big)}{dR}f(n)\Big(\frac{dn}{dR}\Big)\Big]. \label{KSmodel}
\end{align}
\Eq{EL-chi2} is the Schr{\"o}dinger equation for the nuclear wave function, while \Eq{EL-n2} is a differential equation for the electronic density.  The geometric scalar correction is of minor importance in \Eq{EL-chi2}, since it is nonvanishing only in the region where $\chi$ is small. Therefore, the solution $\chi$ of the coupled \Eqs{EL-chi2}--\eqref{EL-n2} is similar to both the one from the BO approximation $\chi^{\rm BO}$ and the exact one.  In \Eq{EL-n2}, however, the nonadiabatic correction due to the nuclear-electronic coupling is expected to play a nontrivial role.  This is mainly through the last term in $v_{\rm geo}$, which involves the $R$-space derivatives of $\ln |\chi|^2$ and $n$, and couples the information in the nuclear and electronic densities at different $R$ points, hence capturing the major nonadiabatic effect.  This term is $\mathcal{O}(M^{-1/2})$ since the logarithmic derivative of $\chi(R)$ is $\mathcal{O}(M^{1/2})$ (see e.g.~Ref.~\onlinecite{eich2016}).  In the Supplemental Material \cite{Supp}, we show that a simplified $v_{\rm geo}$, keeping only its last term, yields almost the same result as faithfully adopting its full expression in \Eq{KSmodel}, and both versions essentially reproduce the exact density.

To obtain the electronic density, one can also transform \Eq{EL-n2} into KS equations \cite{Supp}. It is worth noticing that the LCDA consists in simply adding $v_{\rm geo}$ to the Kohn-Sham potential from \textit{standard} DFT.

\begin{figure}[t]
\includegraphics[width=\columnwidth]{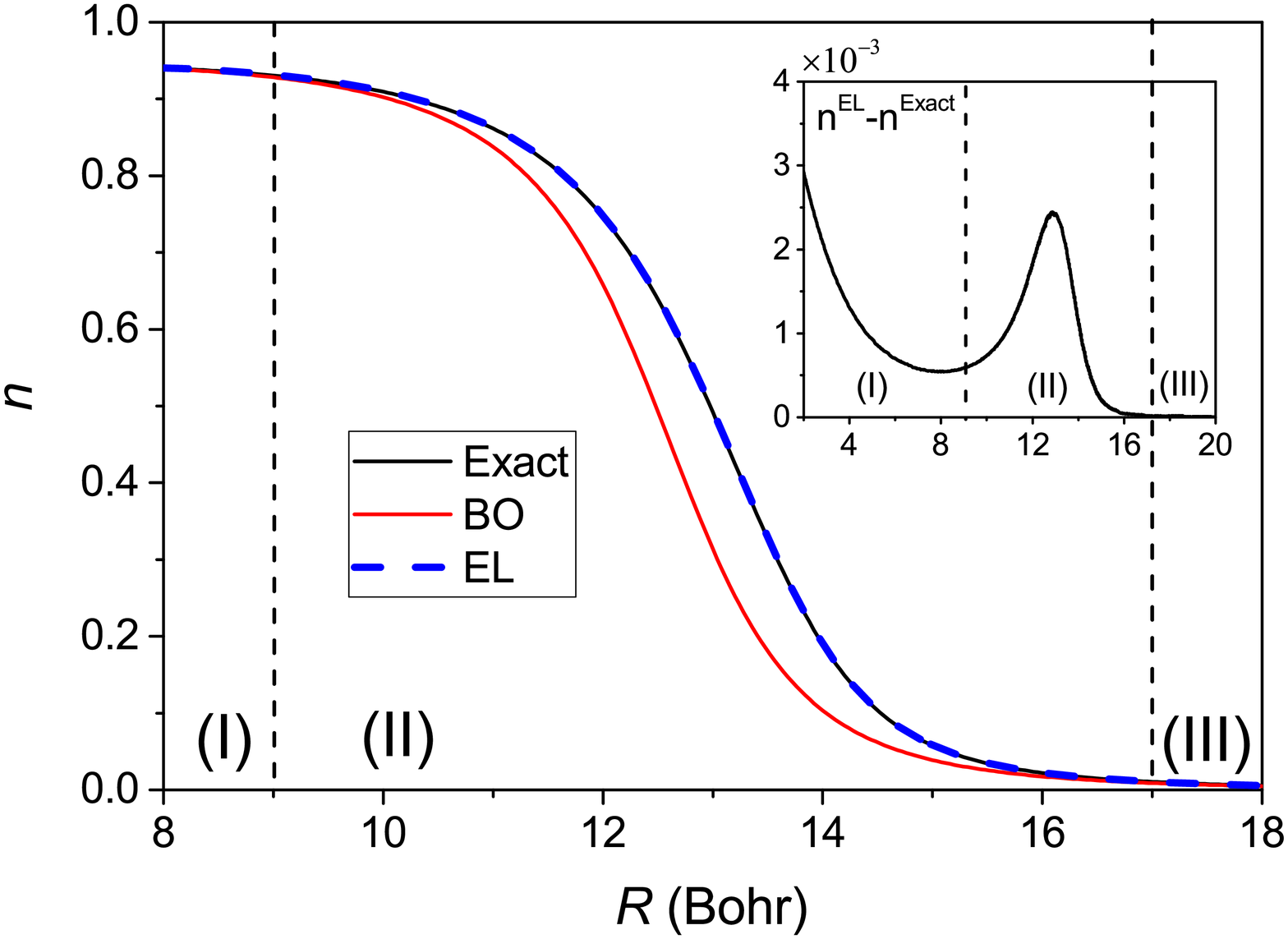}
\caption{
Comparison of the exact and the BO ground state densities.  Also shown is the density obtained by solving the approximate Euler-Lagrange (EL) equation in \Eq{EL-n2}.  \label{fig3}}
\end{figure}

Using the exact nuclear wave function as input, we can solve \Eq{EL-n2} for $n(R)$, and the result is shown in Fig.~\ref{fig3} as the dashed blue curve. As can be seen, the solution of the EL equation almost coincides with the exact density, suggesting that our LCDA is a highly accurate approximation.
Moreover, we note that both curves are close to the BO curve in regions I and III, i.e.~to the left and right of the charge transfer region. This implies that  nonadiabatic effects are small in the $R$-space regions where the density gradient is small. In such regions one can confidently use the BO as a good approximation. Only in region II, where the density is a rapidly changing function of the nuclear configuration, do nonadiabatic effects become nontrivial. However, such regions are probably localized in $R$-space and most likely correspond to charge transfer processes. This reflects that charge transfer is a critical process where nonadiabatic effects are pronounced and the BO approximation might fail qualitatively.

In solving the EL equation or KS equations, one can make use of the localization of nonadiabatic effects and only solve the equation in region II to bridge the BO solutions in region I and III. This should greatly reduce the computational effort.

\section{IV. Exact factorization-based ab initio density functionals for general systems}

With the inspiration gained from our two-site Hubbard model, now we extend this strategy to real systems and formulate $E_e$ as a functional of $\chi(\bdu{R})$ and the continuous conditional electronic density $n_{\bdu{R}}(\br)$.
The discussion is restricted to systems for which the paramagnetic current density $\bm j_p(\br, \bdu{R})$ and the vector potential $\bm A_\mu(\bdu{R})$ are zero. We define $E_e$ through the following constrained search over $\Phi_{\bdu{R}}$:
\begin{align}
    &\quad E_e[\chi,n_{\bdu{R}}] \nonumber\\
    &= \min_{\Phi_{\bdu{R}}\rightarrow n_{\bdu{R}}} \int \Big[\mathcal E_{\rm BO}[\Phi_{\bdu{R}}] + \mathcal E_{\rm geo}[\Phi_{\bdu{R}}]\Big] |\chi(\bdu{R})|^2 d\bdu{R}, \label{Ee-1}
\end{align}
where $\mathcal E_{\rm BO}[\Phi_{\bdu{R}}] = \langle \Phi_{\bdu{R}}|\hat H^{\rm BO}|\Phi_{\bdu{R}}\rangle$
and $\mathcal E_{\rm geo}[\Phi_{\bdu{R}}]$ is given by \Eq{Egeo}.
For practical calculations, we decompose $E_e$ into two parts,
\begin{equation}
    E_e[\chi,n_{\bdu{R}}] = \int E^{\rm BO}[n_{\bdu{R}}]|\chi(\bdu{R})|^2 d\bdu{R} + E_{\rm geo}[\chi, n_{\bdu{R}}], \label{Ee-2}
\end{equation}
where $E^{\rm BO}[n_{\bdu{R}}]$ is the exact density functional in the BO approximation, and $E_{\rm geo}$ is the geometric correction, defined by subtracting the first term on the rhs of \Eq{Ee-2} from $E_e[\chi,n_{\bdu{R}}]$.
Now our local conditional density approximation amounts to approximating $E_{\rm geo}$ as
\begin{equation}
    E_{\rm geo}[\chi, n_{\bdu{R}}] = \int Y[n_{\bdu{R}}, \partial_\mu n_{\bdu{R}}]|\chi(\bdu{R})|^2 d\bdu{R}, \label{LCDA-1}
\end{equation}
where $Y$ is an explicit functional of $n_{\bdu{R}}$ and its gradients $\partial_\mu n_{\bdu{R}}$, in particular,
\begin{equation}
    Y[n_{\bdu{R}}, \partial_\mu n_{\bdu{R}}] = \int \frac{1}{2}y\big(n_{\bdu{R}}(\br)\big) I_{\mu\nu} \partial_\mu n_{\bdu{R}}(\br)\partial_\nu n_{\bdu{R}}(\br) d\br, \label{LCDA-2}
\end{equation}
where $y$ is a local function of $n_{\bdu{R}}(\br)$.

In the one electron case, $\Phi_{\bdu{R}}=\sqrt {n_{\bdu{R}}}$, and one can show that $y$ is given by
\begin{equation}
    y(n_{\bdu{R}}) = \frac{1}{4n_{\bdu{R}}}.
\end{equation}
In the generic many-electron case and under the assumption $\bm j_p(\br, \bdu{R})=\bm A_\mu(\bdu{R})=0$, our LCDA reduces to finding an approximation to the single function $y(n_{\bdu{R}})$.

Alternatively, replacing $\Phi_{\bdu{R}}$ by the KS determinant $\Phi_{\bdu{R}}^{\rm KS}$ in Eq.~(\ref{Egeo}) yields an implicit density functional
\begin{equation}
Y^{\rm KS}[n_{\bdu{R}}, \partial_\mu n_{\bdu{R}}] = \frac{1}{2} I_{\mu\nu} \langle \partial_{\mu} \Phi_{\bdu{R}}^{\rm KS} | \partial_{\nu} \Phi_{\bdu{R}}^{\rm KS} \rangle {.}
\end{equation}

For practical calculations, one can further apply the Kohn-Sham scheme to the electronic part of the problem, i.e., for each $\bdu{R}$, one assumes that $n_{\bdu{R}}(\br)$ comes from a Slater determinant $\det(\psi^1_{\bdu{R}}(\br), \psi^2_{\bdu{R}}(\br), \cdots, \psi^N_{\bdu{R}}(\br))$ and decomposes $E^{\rm BO}$ into the noninteracting electronic kinetic energy $T_{e,s}$, the static Coulomb interaction energies $V_{en}$ and $V_{ee}$, and the exchange-correlation energy $E_{xc}$, $E^{\rm BO} = T_{e,s}+V_{en}+V_{ee}+E_{xc}$.
Similar to \Eqs{EL-chi2}--\eqref{EL-n2}, one can deduce the EL equation for the nuclear wave function and the KS equations for the $\psi^i_{\bdu{R}}(\br)$ as
\begin{align}
    -\frac{1}{2}I_{\mu \nu} \partial^2_{\mu\nu}\chi + \Big(E^{\rm BO}[n_{\bdu{R}}]+Y[n_{\bdu{R}}, \partial_\mu n_{\bdu{R}}]\Big)\chi &= E\chi, \label{nucl} \\
    \Big[-\frac{1}{2}\nabla_{\br}^2  + v_s^{\rm BO}({\br, \bdu{R}})+ v_{\rm geo}(\br,\bdu{R})\Big]\psi^k_{\bdu{R}} &= \lambda^k_{\bdu{R}} \psi^k_{\bdu{R}}. \label{elec}
\end{align}
Here $v_s^{\rm BO}({\br, \bdu{R}})$ is the conventional KS potential in the BO approximation, but with explicit $\bdu{R}$ dependence, i.e.~$v_s^{\rm BO}({\br, \bdu{R}}) = \delta (V_{en}+V_{ee}+E_{xc})/\delta n_{\bdu{R}}$ and $v_{\rm geo}({\br, \bdu{R}})$ is the geometric correction to the potential, given by
\begin{equation}
    v_{\rm geo}({\br,\bdu{R}}) = \frac{\delta Y}{\delta n_{\bdu{R}}} - \frac{1}{|\chi(\bdu{R})|^2}\frac{\partial}{\partial_\mu}\Big(|\chi({\bdu{R}})|^2\frac{\delta Y}{\delta \partial_\mu n_{\bdu{R}}}\Big),
\end{equation}
which after simplification reads
\begin{align}
    v_{\rm geo}(\br, \bdu{R}) &= -\frac{1}{2} I_{\mu\nu}y[n_{\bdu{R}}]\Big\{\partial_\mu + \frac{\partial }{\partial_\mu} \Big(\ln |\chi|^2\Big)\Big\}\partial_\nu n_{\bdu{R}}(\br).
\end{align}
On the rhs of \Eq{elec}, $\lambda_{\bdu{R}}^k$ is the Lagrange multiplier for the normalization constraint on each $\psi_{\bdu{R}}^k$.

The equations in \eqref{elec} are Kohn-Sham equations that take nonadiabatic effects into account. Instead of having a set of independent Kohn-Sham equations for each $\bdu{R}$, we now have a coupled set of equations. One can solve them iteratively together with the nuclear Schr{\"o}dinger equation \eqref{nucl} until self-consistency is reached. In the previous section, we have solved a similar equation, \Eq{EL-n2}, which couples different $R$ points. Alternatively, we can solve the Kohn-Sham equations with the nonadiabatic correction to the KS potential, \Eq{KSmodel}, which is completely equivalent to solving \Eq{EL-n2} directly \cite{Supp}. Therefore, we have carried out the first Kohn-Sham equation with seamless nonadiabatic coupling corrections.

\section{V. Concluding Remarks}

In this work, we have used the asymmetric two-site Hubbard model with $R$-dependence to model a sudden, charge transfer-induced change in the electronic distribution of the ground state conditional electronic wave function of LiF.
By studying nonadiabatic effects, we find that the BO approximation underestimates the critical charge transfer bond length by about 0.5 Bohr. Furthermore, we show that this effect can be perfectly captured in an exact factorization based density functional theory calculation with our newly proposed local conditional density approximation, which leads to coupled equations for the nuclear wave function and conditional electronic density.
This theory is formally exact and in practice reduces the problem to finding a functional approximation for the geometric contribution to the energy expressed in terms of the conditional electronic density $n(\mathbf{r},\bdu{R})$ and nuclear wavefunction $\chi(\bdu{R})$.

Compared with our previously proposed density functional formulation in Ref.~\onlinecite{requist2016b}, we have eliminated the explicit functional dependence on $\mathcal T_{\mu\nu}(\bdu{R})$ [which reduces to a scalar function $g(R)$ in the two-site Hubbard model] in favor of the density and thus greatly simplified the minimization problem. For the general case involving continuous electronic densities, this enables one to seamlessly incorporate beyond-BO effects with only minor modifications to the well-established Kohn-Sham equations without changing its overall framework. 
Thus, the present formulation is an important step towards exact factorization-based ab initio calculations for real applications.

Besides the nonadiabatic correction in the static case, it is reasonable to expect that the additional geometric correction term to the Kohn-Sham potential should play a nontrivial role in dynamical charge transfer processes. Furthermore, since the correction has a prefactor of $1/M$, the nonadiabatic effect should be more pronounced for lighter nuclei, such as in molecules with hydrogen atoms, or in proton coupled charge transfer.  This will involve the time dependent extension of the present theory, which is left for future work.

\bibliography{Reference}

\begin{thebibliography}{55}%
\makeatletter
\providecommand \@ifxundefined [1]{%
 \@ifx{#1\undefined}
}%
\providecommand \@ifnum [1]{%
 \ifnum #1\expandafter \@firstoftwo
 \else \expandafter \@secondoftwo
 \fi
}%
\providecommand \@ifx [1]{%
 \ifx #1\expandafter \@firstoftwo
 \else \expandafter \@secondoftwo
 \fi
}%
\providecommand \natexlab [1]{#1}%
\providecommand \enquote  [1]{``#1''}%
\providecommand \bibnamefont  [1]{#1}%
\providecommand \bibfnamefont [1]{#1}%
\providecommand \citenamefont [1]{#1}%
\providecommand \href@noop [0]{\@secondoftwo}%
\providecommand \href [0]{\begingroup \@sanitize@url \@href}%
\providecommand \@href[1]{\@@startlink{#1}\@@href}%
\providecommand \@@href[1]{\endgroup#1\@@endlink}%
\providecommand \@sanitize@url [0]{\catcode `\\12\catcode `\$12\catcode
  `\&12\catcode `\#12\catcode `\^12\catcode `\_12\catcode `\%12\relax}%
\providecommand \@@startlink[1]{}%
\providecommand \@@endlink[0]{}%
\providecommand \url  [0]{\begingroup\@sanitize@url \@url }%
\providecommand \@url [1]{\endgroup\@href {#1}{\urlprefix }}%
\providecommand \urlprefix  [0]{URL }%
\providecommand \Eprint [0]{\href }%
\providecommand \doibase [0]{http://dx.doi.org/}%
\providecommand \selectlanguage [0]{\@gobble}%
\providecommand \bibinfo  [0]{\@secondoftwo}%
\providecommand \bibfield  [0]{\@secondoftwo}%
\providecommand \translation [1]{[#1]}%
\providecommand \BibitemOpen [0]{}%
\providecommand \bibitemStop [0]{}%
\providecommand \bibitemNoStop [0]{.\EOS\space}%
\providecommand \EOS [0]{\spacefactor3000\relax}%
\providecommand \BibitemShut  [1]{\csname bibitem#1\endcsname}%
\let\auto@bib@innerbib\@empty
%</preamble>
\bibitem [{\citenamefont {Born}\ and\ \citenamefont
  {Oppenheimer}(1927)}]{born1927}%
  \BibitemOpen
  \bibfield  {author} {\bibinfo {author} {\bibfnamefont {M.}~\bibnamefont
  {Born}}\ and\ \bibinfo {author} {\bibfnamefont {R.}~\bibnamefont
  {Oppenheimer}},\ }\href@noop {} {\bibfield  {journal} {\bibinfo  {journal}
  {Ann. Phys.}\ }\textbf {\bibinfo {volume} {84}},\ \bibinfo {pages} {457}
  (\bibinfo {year} {1927})}\BibitemShut {NoStop}%
\bibitem [{\citenamefont {Born}\ and\ \citenamefont {Huang}(1954)}]{born1954}%
  \BibitemOpen
  \bibfield  {author} {\bibinfo {author} {\bibfnamefont {M.}~\bibnamefont
  {Born}}\ and\ \bibinfo {author} {\bibfnamefont {K.}~\bibnamefont {Huang}},\
  }\href@noop {} {\emph {\bibinfo {title} {Dynamical theory of crystal
  lattices}}}\ (\bibinfo  {publisher} {Oxford University Press, New York},\
  \bibinfo {year} {1954})\BibitemShut {NoStop}%
\bibitem [{\citenamefont {Domcke}\ \emph {et~al.}(2011)\citenamefont {Domcke},
  \citenamefont {Yarkony},\ and\ \citenamefont {K{\"o}ppel}}]{domcke2011}%
  \BibitemOpen
  \bibfield  {author} {\bibinfo {author} {\bibfnamefont {W.}~\bibnamefont
  {Domcke}}, \bibinfo {author} {\bibfnamefont {E.}~\bibnamefont {Yarkony}}, \
  and\ \bibinfo {author} {\bibfnamefont {H.}~\bibnamefont {K{\"o}ppel}},\
  }\href@noop {} {\emph {\bibinfo {title} {Conical Intersections, Theory,
  Computation and Experiment}}},\ Vol.~\bibinfo {volume} {17}\ (\bibinfo
  {publisher} {World Scientific, Singapore},\ \bibinfo {year}
  {2011})\BibitemShut {NoStop}%
\bibitem [{\citenamefont {Chen}\ and\ \citenamefont
  {Schaefer}(1980)}]{chen1980}%
  \BibitemOpen
  \bibfield  {author} {\bibinfo {author} {\bibfnamefont {M.~M.~L.}\
  \bibnamefont {Chen}}\ and\ \bibinfo {author} {\bibfnamefont {H.~F.}\
  \bibnamefont {Schaefer}},\ }\href@noop {} {\bibfield  {journal} {\bibinfo
  {journal} {J. Chem. Phys.}\ }\textbf {\bibinfo {volume} {72}},\ \bibinfo
  {pages} {4376} (\bibinfo {year} {1980})}\BibitemShut {NoStop}%
\bibitem [{\citenamefont {Aguado}\ \emph {et~al.}(1995)\citenamefont {Aguado},
  \citenamefont {Su\'arez},\ and\ \citenamefont {Paniagua}}]{aguado1995}%
  \BibitemOpen
  \bibfield  {author} {\bibinfo {author} {\bibfnamefont {A.}~\bibnamefont
  {Aguado}}, \bibinfo {author} {\bibfnamefont {C.}~\bibnamefont {Su\'arez}}, \
  and\ \bibinfo {author} {\bibfnamefont {M.}~\bibnamefont {Paniagua}},\
  }\href@noop {} {\bibfield  {journal} {\bibinfo  {journal} {Chem. Phys.}\
  }\textbf {\bibinfo {volume} {201}},\ \bibinfo {pages} {107} (\bibinfo {year}
  {1995})}\BibitemShut {NoStop}%
\bibitem [{\citenamefont {Ventura}(1996)}]{ventura1996}%
  \BibitemOpen
  \bibfield  {author} {\bibinfo {author} {\bibfnamefont {O.~N.}\ \bibnamefont
  {Ventura}},\ }\href@noop {} {\bibfield  {journal} {\bibinfo  {journal}
  {Molec. Phys.}\ }\textbf {\bibinfo {volume} {89}},\ \bibinfo {pages} {1851}
  (\bibinfo {year} {1996})}\BibitemShut {NoStop}%
\bibitem [{\citenamefont {{Q. Fan, \textit{et al.}}}(2013)}]{fan2013}%
  \BibitemOpen
  \bibfield  {author} {\bibinfo {author} {\bibnamefont {{Q. Fan, \textit{et
  al.}}}},\ }\href@noop {} {\bibfield  {journal} {\bibinfo  {journal} {J. Phys.
  Chem. A}\ }\textbf {\bibinfo {volume} {117}},\ \bibinfo {pages} {10027}
  (\bibinfo {year} {2013})}\BibitemShut {NoStop}%
\bibitem [{\citenamefont {Moutinho}\ \emph {et~al.}(1971)\citenamefont
  {Moutinho}, \citenamefont {Aten},\ and\ \citenamefont {Los}}]{moutinho1971}%
  \BibitemOpen
  \bibfield  {author} {\bibinfo {author} {\bibfnamefont {A.~M.~C.}\
  \bibnamefont {Moutinho}}, \bibinfo {author} {\bibfnamefont {J.~A.}\
  \bibnamefont {Aten}}, \ and\ \bibinfo {author} {\bibfnamefont
  {J.}~\bibnamefont {Los}},\ }\href@noop {} {\bibfield  {journal} {\bibinfo
  {journal} {Physica}\ }\textbf {\bibinfo {volume} {53}},\ \bibinfo {pages}
  {471} (\bibinfo {year} {1971})}\BibitemShut {NoStop}%
\bibitem [{\citenamefont {Faist}\ and\ \citenamefont
  {Levine}(1976)}]{faist1976}%
  \BibitemOpen
  \bibfield  {author} {\bibinfo {author} {\bibfnamefont {M.~B.}\ \bibnamefont
  {Faist}}\ and\ \bibinfo {author} {\bibfnamefont {R.~D.}\ \bibnamefont
  {Levine}},\ }\href@noop {} {\bibfield  {journal} {\bibinfo  {journal} {J.
  Chem. Phys.}\ }\textbf {\bibinfo {volume} {64}},\ \bibinfo {pages} {2953}
  (\bibinfo {year} {1976})}\BibitemShut {NoStop}%
\bibitem [{\citenamefont {Kleyn}\ \emph {et~al.}(1982)\citenamefont {Kleyn},
  \citenamefont {Los},\ and\ \citenamefont {Gislason}}]{kleyn1982}%
  \BibitemOpen
  \bibfield  {author} {\bibinfo {author} {\bibfnamefont {A.~W.}\ \bibnamefont
  {Kleyn}}, \bibinfo {author} {\bibfnamefont {J.}~\bibnamefont {Los}}, \ and\
  \bibinfo {author} {\bibfnamefont {E.~A.}\ \bibnamefont {Gislason}},\
  }\href@noop {} {\bibfield  {journal} {\bibinfo  {journal} {Phys. Rep.}\
  }\textbf {\bibinfo {volume} {90}},\ \bibinfo {pages} {1} (\bibinfo {year}
  {1982})}\BibitemShut {NoStop}%
\bibitem [{\citenamefont {Tully}(1974)}]{tully1974}%
  \BibitemOpen
  \bibfield  {author} {\bibinfo {author} {\bibfnamefont {J.~C.}\ \bibnamefont
  {Tully}},\ }\href@noop {} {\bibfield  {journal} {\bibinfo  {journal} {J.
  Chem. Phys.}\ }\textbf {\bibinfo {volume} {60}},\ \bibinfo {pages} {3042}
  (\bibinfo {year} {1974})}\BibitemShut {NoStop}%
\bibitem [{\citenamefont {Alexander}\ \emph {et~al.}(2000)\citenamefont
  {Alexander}, \citenamefont {Manolopoulos},\ and\ \citenamefont
  {Werner}}]{alexander2000}%
  \BibitemOpen
  \bibfield  {author} {\bibinfo {author} {\bibfnamefont {M.~H.}\ \bibnamefont
  {Alexander}}, \bibinfo {author} {\bibfnamefont {D.~E.}\ \bibnamefont
  {Manolopoulos}}, \ and\ \bibinfo {author} {\bibfnamefont {H.-J.}\
  \bibnamefont {Werner}},\ }\href@noop {} {\bibfield  {journal} {\bibinfo
  {journal} {J. Chem. Phys.}\ }\textbf {\bibinfo {volume} {113}},\ \bibinfo
  {pages} {11084} (\bibinfo {year} {2000})}\BibitemShut {NoStop}%
\bibitem [{\citenamefont {{L. Che, {\it et al.}}}(2007)}]{che2007}%
  \BibitemOpen
  \bibfield  {author} {\bibinfo {author} {\bibnamefont {{L. Che, {\it et
  al.}}}},\ }\href@noop {} {\bibfield  {journal} {\bibinfo  {journal}
  {Science}\ }\textbf {\bibinfo {volume} {317}},\ \bibinfo {pages} {1061}
  (\bibinfo {year} {2007})}\BibitemShut {NoStop}%
\bibitem [{\citenamefont {Lique}\ \emph {et~al.}(2011)\citenamefont {Lique},
  \citenamefont {Li}, \citenamefont {Werner},\ and\ \citenamefont
  {Alexander}}]{lique2011}%
  \BibitemOpen
  \bibfield  {author} {\bibinfo {author} {\bibfnamefont {F.}~\bibnamefont
  {Lique}}, \bibinfo {author} {\bibfnamefont {G.}~\bibnamefont {Li}}, \bibinfo
  {author} {\bibfnamefont {H.-J.}\ \bibnamefont {Werner}}, \ and\ \bibinfo
  {author} {\bibfnamefont {M.~H.}\ \bibnamefont {Alexander}},\ }\href@noop {}
  {\bibfield  {journal} {\bibinfo  {journal} {J. Chem. Phys.}\ }\textbf
  {\bibinfo {volume} {134}},\ \bibinfo {pages} {231101} (\bibinfo {year}
  {2011})}\BibitemShut {NoStop}%
\bibitem [{\citenamefont {Tuckerman}\ \emph {et~al.}(1995)\citenamefont
  {Tuckerman}, \citenamefont {Laasonen}, \citenamefont {Sprik},\ and\
  \citenamefont {Parrinello}}]{tuckerman1995}%
  \BibitemOpen
  \bibfield  {author} {\bibinfo {author} {\bibfnamefont {M.}~\bibnamefont
  {Tuckerman}}, \bibinfo {author} {\bibfnamefont {K.}~\bibnamefont {Laasonen}},
  \bibinfo {author} {\bibfnamefont {M.}~\bibnamefont {Sprik}}, \ and\ \bibinfo
  {author} {\bibfnamefont {M.}~\bibnamefont {Parrinello}},\ }\href@noop {}
  {\bibfield  {journal} {\bibinfo  {journal} {J. Chem. Phys.}\ }\textbf
  {\bibinfo {volume} {103}},\ \bibinfo {pages} {150} (\bibinfo {year}
  {1995})}\BibitemShut {NoStop}%
\bibitem [{\citenamefont {Cheng}\ \emph {et~al.}(1995)\citenamefont {Cheng},
  \citenamefont {Barnett},\ and\ \citenamefont {Landman}}]{cheng1995}%
  \BibitemOpen
  \bibfield  {author} {\bibinfo {author} {\bibfnamefont {H.-P.}\ \bibnamefont
  {Cheng}}, \bibinfo {author} {\bibfnamefont {R.~N.}\ \bibnamefont {Barnett}},
  \ and\ \bibinfo {author} {\bibfnamefont {U.}~\bibnamefont {Landman}},\
  }\href@noop {} {\bibfield  {journal} {\bibinfo  {journal} {Chem. Phys.
  Lett.}\ }\textbf {\bibinfo {volume} {237}},\ \bibinfo {pages} {161} (\bibinfo
  {year} {1995})}\BibitemShut {NoStop}%
\bibitem [{\citenamefont {Lobaugh}\ and\ \citenamefont
  {Voth}(1996)}]{lobaugh1996}%
  \BibitemOpen
  \bibfield  {author} {\bibinfo {author} {\bibfnamefont {J.}~\bibnamefont
  {Lobaugh}}\ and\ \bibinfo {author} {\bibfnamefont {G.~A.}\ \bibnamefont
  {Voth}},\ }\href@noop {} {\bibfield  {journal} {\bibinfo  {journal} {J. Chem.
  Phys.}\ }\textbf {\bibinfo {volume} {104}},\ \bibinfo {pages} {2056}
  (\bibinfo {year} {1996})}\BibitemShut {NoStop}%
\bibitem [{\citenamefont {Tuckerman}\ \emph {et~al.}(1997)\citenamefont
  {Tuckerman}, \citenamefont {Marx}, \citenamefont {Klein},\ and\ \citenamefont
  {Parrinello}}]{tuckerman1997}%
  \BibitemOpen
  \bibfield  {author} {\bibinfo {author} {\bibfnamefont {M.~E.}\ \bibnamefont
  {Tuckerman}}, \bibinfo {author} {\bibfnamefont {D.}~\bibnamefont {Marx}},
  \bibinfo {author} {\bibfnamefont {M.~L.}\ \bibnamefont {Klein}}, \ and\
  \bibinfo {author} {\bibfnamefont {M.}~\bibnamefont {Parrinello}},\
  }\href@noop {} {\bibfield  {journal} {\bibinfo  {journal} {Science}\ }\textbf
  {\bibinfo {volume} {275}},\ \bibinfo {pages} {817} (\bibinfo {year}
  {1997})}\BibitemShut {NoStop}%
\bibitem [{\citenamefont {Fang}\ and\ \citenamefont
  {Hammes-Schiffer}(1997)}]{fang1997}%
  \BibitemOpen
  \bibfield  {author} {\bibinfo {author} {\bibfnamefont {J.-Y.}\ \bibnamefont
  {Fang}}\ and\ \bibinfo {author} {\bibfnamefont {S.}~\bibnamefont
  {Hammes-Schiffer}},\ }\href@noop {} {\bibfield  {journal} {\bibinfo
  {journal} {J. Chem. Phys.}\ }\textbf {\bibinfo {volume} {107}},\ \bibinfo
  {pages} {8933} (\bibinfo {year} {1997})}\BibitemShut {NoStop}%
\bibitem [{\citenamefont {Decornez}\ \emph {et~al.}(1999)\citenamefont
  {Decornez}, \citenamefont {Drukker},\ and\ \citenamefont
  {Hammes-Schiffer}}]{decornez1999}%
  \BibitemOpen
  \bibfield  {author} {\bibinfo {author} {\bibfnamefont {H.}~\bibnamefont
  {Decornez}}, \bibinfo {author} {\bibfnamefont {K.}~\bibnamefont {Drukker}}, \
  and\ \bibinfo {author} {\bibfnamefont {S.}~\bibnamefont {Hammes-Schiffer}},\
  }\href@noop {} {\bibfield  {journal} {\bibinfo  {journal} {J. Phys. Chem. A}\
  }\textbf {\bibinfo {volume} {103}},\ \bibinfo {pages} {2891} (\bibinfo {year}
  {1999})}\BibitemShut {NoStop}%
\bibitem [{\citenamefont {Marx}\ \emph {et~al.}(2010)\citenamefont {Marx},
  \citenamefont {Chandra},\ and\ \citenamefont {Tuckerman}}]{Marx102174}%
  \BibitemOpen
  \bibfield  {author} {\bibinfo {author} {\bibfnamefont {D.}~\bibnamefont
  {Marx}}, \bibinfo {author} {\bibfnamefont {A.}~\bibnamefont {Chandra}}, \
  and\ \bibinfo {author} {\bibfnamefont {M.~E.}\ \bibnamefont {Tuckerman}},\
  }\href@noop {} {\bibfield  {journal} {\bibinfo  {journal} {Chem. Rev.}\
  }\textbf {\bibinfo {volume} {110}},\ \bibinfo {pages} {2174} (\bibinfo {year}
  {2010})}\BibitemShut {NoStop}%
\bibitem [{\citenamefont {Cao}\ \emph {et~al.}(2010)\citenamefont {Cao},
  \citenamefont {Peng}, \citenamefont {Yan}, \citenamefont {Li}, \citenamefont
  {Li},\ and\ \citenamefont {Voth}}]{cao2010}%
  \BibitemOpen
  \bibfield  {author} {\bibinfo {author} {\bibfnamefont {Z.}~\bibnamefont
  {Cao}}, \bibinfo {author} {\bibfnamefont {Y.}~\bibnamefont {Peng}}, \bibinfo
  {author} {\bibfnamefont {T.}~\bibnamefont {Yan}}, \bibinfo {author}
  {\bibfnamefont {S.}~\bibnamefont {Li}}, \bibinfo {author} {\bibfnamefont
  {A.}~\bibnamefont {Li}}, \ and\ \bibinfo {author} {\bibfnamefont {G.~A.}\
  \bibnamefont {Voth}},\ }\href@noop {} {\bibfield  {journal} {\bibinfo
  {journal} {J. Am. Chem. Soc.}\ }\textbf {\bibinfo {volume} {132}},\ \bibinfo
  {pages} {11395} (\bibinfo {year} {2010})}\BibitemShut {NoStop}%
\bibitem [{\citenamefont {Hassanali}\ \emph {et~al.}(2013)\citenamefont
  {Hassanali}, \citenamefont {Giberti}, \citenamefont {Cuny}, \citenamefont
  {K\"uhne},\ and\ \citenamefont {Parrinello}}]{hassanali2013}%
  \BibitemOpen
  \bibfield  {author} {\bibinfo {author} {\bibfnamefont {A.}~\bibnamefont
  {Hassanali}}, \bibinfo {author} {\bibfnamefont {F.}~\bibnamefont {Giberti}},
  \bibinfo {author} {\bibfnamefont {J.}~\bibnamefont {Cuny}}, \bibinfo {author}
  {\bibfnamefont {T.~D.}\ \bibnamefont {K\"uhne}}, \ and\ \bibinfo {author}
  {\bibfnamefont {M.}~\bibnamefont {Parrinello}},\ }\href@noop {} {\bibfield
  {journal} {\bibinfo  {journal} {PNAS}\ }\textbf {\bibinfo {volume} {110}},\
  \bibinfo {pages} {13723} (\bibinfo {year} {2013})}\BibitemShut {NoStop}%
\bibitem [{\citenamefont {Rossi}\ \emph {et~al.}(2016)\citenamefont {Rossi},
  \citenamefont {Ceriotti},\ and\ \citenamefont {Manolopoulos}}]{rossi2016}%
  \BibitemOpen
  \bibfield  {author} {\bibinfo {author} {\bibfnamefont {M.}~\bibnamefont
  {Rossi}}, \bibinfo {author} {\bibfnamefont {M.}~\bibnamefont {Ceriotti}}, \
  and\ \bibinfo {author} {\bibfnamefont {D.~E.}\ \bibnamefont {Manolopoulos}},\
  }\href@noop {} {\bibfield  {journal} {\bibinfo  {journal} {J. Phys. Chem.
  Lett.}\ }\textbf {\bibinfo {volume} {7}},\ \bibinfo {pages} {3001} (\bibinfo
  {year} {2016})}\BibitemShut {NoStop}%
\bibitem [{\citenamefont {Kreibich}\ and\ \citenamefont
  {Gross}(2001)}]{kreibich2001}%
  \BibitemOpen
  \bibfield  {author} {\bibinfo {author} {\bibfnamefont {T.}~\bibnamefont
  {Kreibich}}\ and\ \bibinfo {author} {\bibfnamefont {E.~K.~U.}\ \bibnamefont
  {Gross}},\ }\href@noop {} {\bibfield  {journal} {\bibinfo  {journal} {Phys.
  Rev. Lett.}\ }\textbf {\bibinfo {volume} {86}},\ \bibinfo {pages} {2984}
  (\bibinfo {year} {2001})}\BibitemShut {NoStop}%
\bibitem [{\citenamefont {Kreibich}\ \emph {et~al.}(2008)\citenamefont
  {Kreibich}, \citenamefont {{R. van Leeuwen}},\ and\ \citenamefont
  {Gross}}]{kreibich2008}%
  \BibitemOpen
  \bibfield  {author} {\bibinfo {author} {\bibfnamefont {T.}~\bibnamefont
  {Kreibich}}, \bibinfo {author} {\bibnamefont {{R. van Leeuwen}}}, \ and\
  \bibinfo {author} {\bibfnamefont {E.~K.~U.}\ \bibnamefont {Gross}},\
  }\href@noop {} {\bibfield  {journal} {\bibinfo  {journal} {Phys. Rev. A}\
  }\textbf {\bibinfo {volume} {78}},\ \bibinfo {pages} {022501} (\bibinfo
  {year} {2008})}\BibitemShut {NoStop}%
\bibitem [{\citenamefont {Chakraborty}\ \emph {et~al.}(2008)\citenamefont
  {Chakraborty}, \citenamefont {Pak},\ and\ \citenamefont
  {Hammes-Schiffer}}]{chakraborty2008}%
  \BibitemOpen
  \bibfield  {author} {\bibinfo {author} {\bibfnamefont {A.}~\bibnamefont
  {Chakraborty}}, \bibinfo {author} {\bibfnamefont {M.~V.}\ \bibnamefont
  {Pak}}, \ and\ \bibinfo {author} {\bibfnamefont {S.}~\bibnamefont
  {Hammes-Schiffer}},\ }\href@noop {} {\bibfield  {journal} {\bibinfo
  {journal} {Phys. Rev. Lett.}\ }\textbf {\bibinfo {volume} {101}},\ \bibinfo
  {pages} {153001} (\bibinfo {year} {2008})}\BibitemShut {NoStop}%
\bibitem [{\citenamefont {Yang}\ \emph {et~al.}(2017)\citenamefont {Yang},
  \citenamefont {Brorsen}, \citenamefont {Culpitt}, \citenamefont {Pak},\ and\
  \citenamefont {Hammes-Schiffer}}]{yang2017}%
  \BibitemOpen
  \bibfield  {author} {\bibinfo {author} {\bibfnamefont {Y.}~\bibnamefont
  {Yang}}, \bibinfo {author} {\bibfnamefont {K.~R.}\ \bibnamefont {Brorsen}},
  \bibinfo {author} {\bibfnamefont {T.}~\bibnamefont {Culpitt}}, \bibinfo
  {author} {\bibfnamefont {M.~V.}\ \bibnamefont {Pak}}, \ and\ \bibinfo
  {author} {\bibfnamefont {S.}~\bibnamefont {Hammes-Schiffer}},\ }\href@noop {}
  {\bibfield  {journal} {\bibinfo  {journal} {J. Chem. Phys.}\ }\textbf
  {\bibinfo {volume} {147}},\ \bibinfo {pages} {114113} (\bibinfo {year}
  {2017})}\BibitemShut {NoStop}%
\bibitem [{\citenamefont {Hunter}(1975{\natexlab{a}})}]{hunter1975}%
  \BibitemOpen
  \bibfield  {author} {\bibinfo {author} {\bibfnamefont {G.}~\bibnamefont
  {Hunter}},\ }\href@noop {} {\bibfield  {journal} {\bibinfo  {journal} {Int.
  J. Quantum Chem.}\ }\textbf {\bibinfo {volume} {9}},\ \bibinfo {pages} {237}
  (\bibinfo {year} {1975}{\natexlab{a}})}\BibitemShut {NoStop}%
\bibitem [{\citenamefont {Gidopoulos}\ and\ \citenamefont
  {Gross}(2014)}]{gidopoulos2014}%
  \BibitemOpen
  \bibfield  {author} {\bibinfo {author} {\bibfnamefont {N.~I.}\ \bibnamefont
  {Gidopoulos}}\ and\ \bibinfo {author} {\bibfnamefont {E.~K.~U.}\ \bibnamefont
  {Gross}},\ }\href@noop {} {\bibfield  {journal} {\bibinfo  {journal} {Philos.
  Trans. R. Soc. Lond., A}\ }\textbf {\bibinfo {volume} {372}},\ \bibinfo
  {pages} {20130059} (\bibinfo {year} {2014})}\BibitemShut {NoStop}%
\bibitem [{\citenamefont {Abedi}\ \emph {et~al.}(2010)\citenamefont {Abedi},
  \citenamefont {Maitra},\ and\ \citenamefont {Gross}}]{abedi2010}%
  \BibitemOpen
  \bibfield  {author} {\bibinfo {author} {\bibfnamefont {A.}~\bibnamefont
  {Abedi}}, \bibinfo {author} {\bibfnamefont {N.~T.}\ \bibnamefont {Maitra}}, \
  and\ \bibinfo {author} {\bibfnamefont {E.~K.~U.}\ \bibnamefont {Gross}},\
  }\href@noop {} {\bibfield  {journal} {\bibinfo  {journal} {Phys. Rev. Lett.}\
  }\textbf {\bibinfo {volume} {105}},\ \bibinfo {pages} {123002} (\bibinfo
  {year} {2010})}\BibitemShut {NoStop}%
\bibitem [{\citenamefont {Requist}\ and\ \citenamefont
  {Gross}(2016)}]{requist2016b}%
  \BibitemOpen
  \bibfield  {author} {\bibinfo {author} {\bibfnamefont {R.}~\bibnamefont
  {Requist}}\ and\ \bibinfo {author} {\bibfnamefont {E.~K.~U.}\ \bibnamefont
  {Gross}},\ }\href@noop {} {\bibfield  {journal} {\bibinfo  {journal} {Phys.
  Rev. Lett.}\ }\textbf {\bibinfo {volume} {117}},\ \bibinfo {pages} {193001}
  (\bibinfo {year} {2016})}\BibitemShut {NoStop}%
\bibitem [{\citenamefont {Kohn}\ and\ \citenamefont {Sham}(1965)}]{Koh65A1133}%
  \BibitemOpen
  \bibfield  {author} {\bibinfo {author} {\bibfnamefont {W.}~\bibnamefont
  {Kohn}}\ and\ \bibinfo {author} {\bibfnamefont {L.~J.}\ \bibnamefont
  {Sham}},\ }\href@noop {} {\bibfield  {journal} {\bibinfo  {journal} {Phys.
  Rev.}\ }\textbf {\bibinfo {volume} {140}},\ \bibinfo {pages} {A1133}
  (\bibinfo {year} {1965})}\BibitemShut {NoStop}%
\bibitem [{\citenamefont {Li}\ \emph {et~al.}(2015)\citenamefont {Li},
  \citenamefont {Zheng}, \citenamefont {Cohen}, \citenamefont
  {Mori-S{\'a}nchez},\ and\ \citenamefont {Yang}}]{Li15053001}%
  \BibitemOpen
  \bibfield  {author} {\bibinfo {author} {\bibfnamefont {C.}~\bibnamefont
  {Li}}, \bibinfo {author} {\bibfnamefont {X.}~\bibnamefont {Zheng}}, \bibinfo
  {author} {\bibfnamefont {A.~J.}\ \bibnamefont {Cohen}}, \bibinfo {author}
  {\bibfnamefont {P.}~\bibnamefont {Mori-S{\'a}nchez}}, \ and\ \bibinfo
  {author} {\bibfnamefont {W.}~\bibnamefont {Yang}},\ }\href@noop {} {\bibfield
   {journal} {\bibinfo  {journal} {Phys. Rev. Lett.}\ }\textbf {\bibinfo
  {volume} {114}},\ \bibinfo {pages} {053001} (\bibinfo {year}
  {2015})}\BibitemShut {NoStop}%
\bibitem [{\citenamefont {Ruzsinszky}\ \emph {et~al.}(2006)\citenamefont
  {Ruzsinszky}, \citenamefont {Perdew}, \citenamefont {Csonka}, \citenamefont
  {Vydrov},\ and\ \citenamefont {Scuseria}}]{Ruzsinszky06194112}%
  \BibitemOpen
  \bibfield  {author} {\bibinfo {author} {\bibfnamefont {A.}~\bibnamefont
  {Ruzsinszky}}, \bibinfo {author} {\bibfnamefont {J.~P.}\ \bibnamefont
  {Perdew}}, \bibinfo {author} {\bibfnamefont {G.~I.}\ \bibnamefont {Csonka}},
  \bibinfo {author} {\bibfnamefont {O.~A.}\ \bibnamefont {Vydrov}}, \ and\
  \bibinfo {author} {\bibfnamefont {G.~E.}\ \bibnamefont {Scuseria}},\
  }\href@noop {} {\bibfield  {journal} {\bibinfo  {journal} {J. Chem. Phys.}\
  }\textbf {\bibinfo {volume} {125}},\ \bibinfo {pages} {194112} (\bibinfo
  {year} {2006})}\BibitemShut {NoStop}%
\bibitem [{\citenamefont {Zewail}(2000)}]{Zewail005660}%
  \BibitemOpen
  \bibfield  {author} {\bibinfo {author} {\bibfnamefont {A.~H.}\ \bibnamefont
  {Zewail}},\ }\href@noop {} {\bibfield  {journal} {\bibinfo  {journal} {J.
  Phys. Chem. A}\ }\textbf {\bibinfo {volume} {104}},\ \bibinfo {pages} {5660}
  (\bibinfo {year} {2000})}\BibitemShut {NoStop}%
\bibitem [{\citenamefont {Shin}\ and\ \citenamefont
  {Metiu}(1995)}]{Shin959285}%
  \BibitemOpen
  \bibfield  {author} {\bibinfo {author} {\bibfnamefont {S.}~\bibnamefont
  {Shin}}\ and\ \bibinfo {author} {\bibfnamefont {H.}~\bibnamefont {Metiu}},\
  }\href@noop {} {\bibfield  {journal} {\bibinfo  {journal} {J. Chem. Phys.}\
  }\textbf {\bibinfo {volume} {102}},\ \bibinfo {pages} {9285} (\bibinfo {year}
  {1995})}\BibitemShut {NoStop}%
\bibitem [{\citenamefont {Abedi}\ \emph {et~al.}(2013)\citenamefont {Abedi},
  \citenamefont {Agostini}, \citenamefont {Suzuki},\ and\ \citenamefont
  {Gross}}]{abedi2013}%
  \BibitemOpen
  \bibfield  {author} {\bibinfo {author} {\bibfnamefont {A.}~\bibnamefont
  {Abedi}}, \bibinfo {author} {\bibfnamefont {F.}~\bibnamefont {Agostini}},
  \bibinfo {author} {\bibfnamefont {Y.}~\bibnamefont {Suzuki}}, \ and\ \bibinfo
  {author} {\bibfnamefont {E.~K.~U.}\ \bibnamefont {Gross}},\ }\href@noop {}
  {\bibfield  {journal} {\bibinfo  {journal} {Phys. Rev. Lett.}\ }\textbf
  {\bibinfo {volume} {110}},\ \bibinfo {pages} {263001} (\bibinfo {year}
  {2013})}\BibitemShut {NoStop}%
\bibitem [{\citenamefont {Agostini}\ \emph {et~al.}(2015)\citenamefont
  {Agostini}, \citenamefont {Abedi}, \citenamefont {Suzuki}, \citenamefont
  {Min}, \citenamefont {Maitra},\ and\ \citenamefont {Gross}}]{agostini2015}%
  \BibitemOpen
  \bibfield  {author} {\bibinfo {author} {\bibfnamefont {F.}~\bibnamefont
  {Agostini}}, \bibinfo {author} {\bibfnamefont {A.}~\bibnamefont {Abedi}},
  \bibinfo {author} {\bibfnamefont {Y.}~\bibnamefont {Suzuki}}, \bibinfo
  {author} {\bibfnamefont {S.~K.}\ \bibnamefont {Min}}, \bibinfo {author}
  {\bibfnamefont {N.~T.}\ \bibnamefont {Maitra}}, \ and\ \bibinfo {author}
  {\bibfnamefont {E.~K.~U.}\ \bibnamefont {Gross}},\ }\href@noop {} {\bibfield
  {journal} {\bibinfo  {journal} {J. Chem. Phys.}\ }\textbf {\bibinfo {volume}
  {142}},\ \bibinfo {pages} {084303} (\bibinfo {year} {2015})}\BibitemShut
  {NoStop}%
\bibitem [{\citenamefont {Min}\ \emph {et~al.}(2017)\citenamefont {Min},
  \citenamefont {Agostini}, \citenamefont {Tavernelli},\ and\ \citenamefont
  {Gross}}]{min2017}%
  \BibitemOpen
  \bibfield  {author} {\bibinfo {author} {\bibfnamefont {S.~K.}\ \bibnamefont
  {Min}}, \bibinfo {author} {\bibfnamefont {F.}~\bibnamefont {Agostini}},
  \bibinfo {author} {\bibfnamefont {I.}~\bibnamefont {Tavernelli}}, \ and\
  \bibinfo {author} {\bibfnamefont {E.~K.~U.}\ \bibnamefont {Gross}},\
  }\href@noop {} {\bibfield  {journal} {\bibinfo  {journal} {J. Phys. Chem.
  Lett.}\ }\textbf {\bibinfo {volume} {8}},\ \bibinfo {pages} {3048} (\bibinfo
  {year} {2017})}\BibitemShut {NoStop}%
\bibitem [{\citenamefont {Hunter}(1975{\natexlab{b}})}]{Hunter75237}%
  \BibitemOpen
  \bibfield  {author} {\bibinfo {author} {\bibfnamefont {G.}~\bibnamefont
  {Hunter}},\ }\href@noop {} {\bibfield  {journal} {\bibinfo  {journal} {Int.
  J. Quantum Chem.}\ }\textbf {\bibinfo {volume} {9}},\ \bibinfo {pages} {237}
  (\bibinfo {year} {1975}{\natexlab{b}})}\BibitemShut {NoStop}%
\bibitem [{\citenamefont {Requist}\ \emph {et~al.}(2016)\citenamefont
  {Requist}, \citenamefont {Tandetzky},\ and\ \citenamefont
  {Gross}}]{Requist16042108}%
  \BibitemOpen
  \bibfield  {author} {\bibinfo {author} {\bibfnamefont {R.}~\bibnamefont
  {Requist}}, \bibinfo {author} {\bibfnamefont {F.}~\bibnamefont {Tandetzky}},
  \ and\ \bibinfo {author} {\bibfnamefont {E.~K.~U.}\ \bibnamefont {Gross}},\
  }\href@noop {} {\bibfield  {journal} {\bibinfo  {journal} {Phys. Rev. A}\
  }\textbf {\bibinfo {volume} {93}},\ \bibinfo {pages} {042108} (\bibinfo
  {year} {2016})}\BibitemShut {NoStop}%
\bibitem [{Sup()}]{Supp}%
  \BibitemOpen
  \href@noop {} {\ }\bibinfo {note} {See the Supplemental Material for
  details.}\BibitemShut {Stop}%
\bibitem [{\citenamefont {C\'{\i}\v{z}ek}(1969)}]{Cizek69}%
  \BibitemOpen
  \bibfield  {author} {\bibinfo {author} {\bibfnamefont {J.}~\bibnamefont
  {C\'{\i}\v{z}ek}},\ }in\ \href@noop {} {\emph {\bibinfo {booktitle} {Advances
  in Chemical Physics}}},\ Vol.~\bibinfo {volume} {14},\ \bibinfo {editor}
  {edited by\ \bibinfo {editor} {\bibfnamefont {P.~C.}\ \bibnamefont
  {Hariharan}}}\ (\bibinfo  {publisher} {Wiley Interscience, New York},\
  \bibinfo {year} {1969})\ p.~\bibinfo {pages} {35}\BibitemShut {NoStop}%
\bibitem [{\citenamefont {{Purvis~III}}\ and\ \citenamefont
  {Bartlett}(1982)}]{Purvis82}%
  \BibitemOpen
  \bibfield  {author} {\bibinfo {author} {\bibfnamefont {G.~D.}\ \bibnamefont
  {{Purvis~III}}}\ and\ \bibinfo {author} {\bibfnamefont {R.~J.}\ \bibnamefont
  {Bartlett}},\ }\href@noop {} {\bibfield  {journal} {\bibinfo  {journal} {J.
  Chem. Phys.}\ }\textbf {\bibinfo {volume} {76}},\ \bibinfo {pages} {1910}
  (\bibinfo {year} {1982})}\BibitemShut {NoStop}%
\bibitem [{\citenamefont {Pople}\ \emph {et~al.}(1987)\citenamefont {Pople},
  \citenamefont {Head-Gordon},\ and\ \citenamefont {Raghavachari}}]{Pople87}%
  \BibitemOpen
  \bibfield  {author} {\bibinfo {author} {\bibfnamefont {J.~A.}\ \bibnamefont
  {Pople}}, \bibinfo {author} {\bibfnamefont {M.}~\bibnamefont {Head-Gordon}},
  \ and\ \bibinfo {author} {\bibfnamefont {K.}~\bibnamefont {Raghavachari}},\
  }\href@noop {} {\bibfield  {journal} {\bibinfo  {journal} {J. Chem. Phys.}\
  }\textbf {\bibinfo {volume} {87}},\ \bibinfo {pages} {5968} (\bibinfo {year}
  {1987})}\BibitemShut {NoStop}%
\bibitem [{\citenamefont {Hofmann}\ \emph {et~al.}(2001)\citenamefont
  {Hofmann}, \citenamefont {Bockstedte},\ and\ \citenamefont
  {Pankratov}}]{Hofmann01245321}%
  \BibitemOpen
  \bibfield  {author} {\bibinfo {author} {\bibfnamefont {M.}~\bibnamefont
  {Hofmann}}, \bibinfo {author} {\bibfnamefont {M.}~\bibnamefont {Bockstedte}},
  \ and\ \bibinfo {author} {\bibfnamefont {O.}~\bibnamefont {Pankratov}},\
  }\href@noop {} {\bibfield  {journal} {\bibinfo  {journal} {Phys. Rev. B}\
  }\textbf {\bibinfo {volume} {64}},\ \bibinfo {pages} {245321} (\bibinfo
  {year} {2001})}\BibitemShut {NoStop}%
\bibitem [{\citenamefont {Bauschlicher}\ and\ \citenamefont
  {Langhoff}(1988)}]{bauschlicher1988}%
  \BibitemOpen
  \bibfield  {author} {\bibinfo {author} {\bibfnamefont {C.~W.}\ \bibnamefont
  {Bauschlicher}}\ and\ \bibinfo {author} {\bibfnamefont {S.~R.}\ \bibnamefont
  {Langhoff}},\ }\href@noop {} {\bibfield  {journal} {\bibinfo  {journal} {J.
  Chem. Phys.}\ }\textbf {\bibinfo {volume} {89}},\ \bibinfo {pages} {4246}
  (\bibinfo {year} {1988})}\BibitemShut {NoStop}%
\bibitem [{\citenamefont {Hanrath}(2008)}]{hanrath2008}%
  \BibitemOpen
  \bibfield  {author} {\bibinfo {author} {\bibfnamefont {M.}~\bibnamefont
  {Hanrath}},\ }\href@noop {} {\bibfield  {journal} {\bibinfo  {journal}
  {Molec. Phys.}\ }\textbf {\bibinfo {volume} {106}},\ \bibinfo {pages} {1949}
  (\bibinfo {year} {2008})}\BibitemShut {NoStop}%
\bibitem [{\citenamefont {Weck}\ \emph {et~al.}(2004)\citenamefont {Weck},
  \citenamefont {Kirby},\ and\ \citenamefont {Stancil}}]{weck2004}%
  \BibitemOpen
  \bibfield  {author} {\bibinfo {author} {\bibfnamefont {P.~F.}\ \bibnamefont
  {Weck}}, \bibinfo {author} {\bibfnamefont {K.}~\bibnamefont {Kirby}}, \ and\
  \bibinfo {author} {\bibfnamefont {P.~C.}\ \bibnamefont {Stancil}},\
  }\href@noop {} {\bibfield  {journal} {\bibinfo  {journal} {J. Chem. Phys.}\
  }\textbf {\bibinfo {volume} {120}},\ \bibinfo {pages} {4216} (\bibinfo {year}
  {2004})}\BibitemShut {NoStop}%
\bibitem [{\citenamefont {Kurosaki}\ and\ \citenamefont
  {Yokoyama}(2012)}]{kurosaki2012}%
  \BibitemOpen
  \bibfield  {author} {\bibinfo {author} {\bibfnamefont {Y.}~\bibnamefont
  {Kurosaki}}\ and\ \bibinfo {author} {\bibfnamefont {K.}~\bibnamefont
  {Yokoyama}},\ }\href@noop {} {\bibfield  {journal} {\bibinfo  {journal} {J.
  Chem. Phys.}\ }\textbf {\bibinfo {volume} {137}},\ \bibinfo {pages} {064305}
  (\bibinfo {year} {2012})}\BibitemShut {NoStop}%
\bibitem [{\citenamefont {Sch\"onhammer}\ \emph {et~al.}(1995)\citenamefont
  {Sch\"onhammer}, \citenamefont {Gunnarsson},\ and\ \citenamefont
  {Noack}}]{schoenhammer1995}%
  \BibitemOpen
  \bibfield  {author} {\bibinfo {author} {\bibfnamefont {K.}~\bibnamefont
  {Sch\"onhammer}}, \bibinfo {author} {\bibfnamefont {O.}~\bibnamefont
  {Gunnarsson}}, \ and\ \bibinfo {author} {\bibfnamefont {R.~M.}\ \bibnamefont
  {Noack}},\ }\href@noop {} {\bibfield  {journal} {\bibinfo  {journal} {Phys.
  Rev. B}\ }\textbf {\bibinfo {volume} {52}},\ \bibinfo {pages} {2504}
  (\bibinfo {year} {1995})}\BibitemShut {NoStop}%
\bibitem [{\citenamefont {Carrascal}\ \emph {et~al.}(2015)\citenamefont
  {Carrascal}, \citenamefont {Ferrer}, \citenamefont {Smith},\ and\
  \citenamefont {Burke}}]{Carrascal15393001}%
  \BibitemOpen
  \bibfield  {author} {\bibinfo {author} {\bibfnamefont {D.~J.}\ \bibnamefont
  {Carrascal}}, \bibinfo {author} {\bibfnamefont {J.}~\bibnamefont {Ferrer}},
  \bibinfo {author} {\bibfnamefont {J.~C.}\ \bibnamefont {Smith}}, \ and\
  \bibinfo {author} {\bibfnamefont {K.}~\bibnamefont {Burke}},\ }\href@noop {}
  {\bibfield  {journal} {\bibinfo  {journal} {J. Phys. Condens. Matter}\
  }\textbf {\bibinfo {volume} {27}},\ \bibinfo {pages} {393001} (\bibinfo
  {year} {2015})}\BibitemShut {NoStop}%
\bibitem [{\citenamefont {Levy}(1979)}]{Levy796062}%
  \BibitemOpen
  \bibfield  {author} {\bibinfo {author} {\bibfnamefont {M.}~\bibnamefont
  {Levy}},\ }\href@noop {} {\bibfield  {journal} {\bibinfo  {journal} {Proc.
  Nat. Acad. Sci}\ }\textbf {\bibinfo {volume} {76}},\ \bibinfo {pages} {6062}
  (\bibinfo {year} {1979})}\BibitemShut {NoStop}%
\bibitem [{\citenamefont {Eich}\ and\ \citenamefont
  {Agostini}(2016)}]{eich2016}%
  \BibitemOpen
  \bibfield  {author} {\bibinfo {author} {\bibfnamefont {F.}~\bibnamefont
  {Eich}}\ and\ \bibinfo {author} {\bibfnamefont {F.}~\bibnamefont
  {Agostini}},\ }\href@noop {} {\bibfield  {journal} {\bibinfo  {journal} {J.
  Chem. Phys.}\ }\textbf {\bibinfo {volume} {145}},\ \bibinfo {pages} {054110}
  (\bibinfo {year} {2016})}\BibitemShut {NoStop}%
\end{thebibliography}%

\end{document}

\draft

Supporting Information for
%
%\vspace{0.2cm}
\vspace{0.1cm}

\begin{center}
{\large \textbf{Exact Factorization Based Density Functional Study of Non-adiabatic Effects in Charge Transfers through 2-Site Hubbard Model}}
\vspace{0.2cm}

Chen Li$^1$, Ryan Requist$^1$ and E. K. U. Gross$^1$$^,$$^2$

{\small
\emph{$^1$Max Planck Institute of Microstructure Physics, Weinberg 2, 06120, Halle, Germany
\\}
\emph{$^2$Fritz Haber Center for Molecular Dynamics, Institute of Chemistry, The Hebrew University of Jerusalem, Jerusalem 91904 Israel
\\}
}

\vspace{0.5cm}
%\today

(Dated: \today)

\end{center}

\linespread{1.0}

%\vspace{0.2cm}

\tableofcontents

\clearpage

%%%%%%%%%%%%%%%%%%%
\renewcommand{\theequation}{S\arabic{equation}}
\renewcommand{\thetable}{S\arabic{table}}
\renewcommand{\thefigure}{S\arabic{figure}}
%%%%%%%%%%%%%%%%%%%%%

\section{Some details about the 2-site Hubbard model}

\subsection{$R$-dependence of $\Delta \ep$ in the Hubbard model}

$\Delta \ep$ is the on-site energy difference. When $R\rightarrow \infty$, $\ep_i \rightarrow -I^i$. Here we derive the asymptotic behavior of $\Delta \ep$ for large $R$. At finite $R$, the electron at the Li atom (site 1) feels the Coulomb interaction from the distant F nucleus and the electron nearby which occupies the $3p_z$ orbital (here we assume the bond axis is the z axis). Note that the effective charge of the F nucleus is +1 after compensating the non-valence electrons. Therefore, $\ep_1$ can be approximately written as
\begin{equation}
    \ep_1 = -I^1 - \int d\br \frac{|\phi_1(\br)|^2}{|\br-\bR_2|} + \iint d\br_1 d\br_2 \frac{|\phi_1(\br_1)|^2 |\phi_2(\br_2)|^2}{|\br_1-\br_2|}.
\end{equation}
Here $\phi_1(\br)=\phi_s(\br-\bR_1)$, and $\phi_2(\br)=\phi_p(\br-\bR_2)$, where $\phi_s$ is the $2s$ orbital of Li and $\phi_p$ is the $3p_z$ orbital of F. $\phi_s$ and $\phi_p$ are both centered at $\br=0$. Similarly, $\ep_2$ can be written as,
\begin{equation}
    \ep_2 = -I^2 - \int d\br \frac{|\phi_2(\br)|^2}{|\br-\bR_1|} + \iint d\br_1 d\br_2 \frac{|\phi_1(\br_1)|^2 |\phi_2(\br_2)|^2}{|\br_1-\br_2|}.
\end{equation}
It follows that
\begin{align}
    \Delta \ep &= I^2-I^1 + \int d\br \frac{|\phi_2(\br)|^2}{|\br-\bR_1|}- \int d\br \frac{|\phi_1(\br)|^2}{|\br-\bR_2|} \nonumber \\
    &=\Delta I + \Delta\ep^1(R),
\end{align}
where $\Delta I = I^2 - I^1$, and
\begin{align}
    \Delta\ep^1(R) &= \int d\br \frac{|\phi_2(\br)|^2}{|\br-\bR_1|}- \int d\br \frac{|\phi_1(\br)|^2}{|\br-\bR_2|} \nonumber \\
    &= \int d\br \frac{|\phi_p(\br-\bR_2)|^2}{|\br-\bR_1|}- \int d\br \frac{|\phi_s(\br-\bR_1)|^2}{|\br-\bR_2|} \nonumber \\
    &= \int d\br \frac{|\phi_p(\br)|^2}{|\br+\bR_{12}|}- \int d\br \frac{|\phi_s(\br)|^2}{|\br-\bR_{12}|}. \label{Dep-1}
\end{align}
Here $\bR_{12} \equiv \bR_2-\bR_1 \equiv R\bf e_z$. By the multipole expansion up to the second order,
\begin{equation}
    \frac{1}{|\br\mp R\bf e_z|} = \frac{1}{R} \pm \frac{\bf e_z \cdot \br}{R^2} + \frac{r^2}{2R^3}[3({\bf e_z \cdot \bf e_{\br})}^2-1]. \label{multipole}
\end{equation}
Substituting \Eq{multipole} into \Eq{Dep-1}, we have
\begin{align}
    \Delta\ep^1(R) &= \int d\br |\phi_p(\br)|^2 \Big\{ \frac{1}{R} - \frac{\bf e_z \cdot \br}{R^2} + \frac{r^2}{2R^3}[3({\bf e_z \cdot \bf e_{\br})}^2-1] \Big\} \nonumber \\
    &\quad - \int d\br |\phi_s(\br)|^2\Big\{ \frac{1}{R} + \frac{\bf e_z \cdot \br}{R^2} + \frac{r^2}{2R^3}[3({\bf e_z \cdot \bf e_{\br})}^2-1]\Big\}.
\end{align}
The $\frac{1}{R}$ term is canceled out since $\phi_s$ and $\phi_p$ both integrate to constant 1. The $\frac{1}{R^2}$ term also vanishes because $|\phi_q|^2\;(q=s,p)$ is an even function of $z$ and hence $\int d\br |\phi_q(\br)|^2 {\bf {e_z} \cdot \br} = \int d\br z|\phi_q(\br)|^2=0$. Therefore only $\frac{1}{R^3}$ term survives and $\Delta\ep^1 (R)$ reads
\begin{align}
    \Delta\ep^1(R) &= \frac{\gamma}{R^3}, \label{Dep-2}
\end{align}
where
\begin{equation}
    \gamma = \int d\br \;r^2 \Big[|\phi_p(\br)|^2 - |\phi_s(\br)|^2\Big] \Big[3(\bf e_z \cdot \bf e_{\br})^2-1\Big].
\end{equation}
We note that \Eq{Dep-2} captures the leading order term in the asymptotic limit of $R\rightarrow \infty$, but is not valid at the limit of $R \rightarrow 0$. To use a simple form to accommodate both limits, we modify \Eq{Dep-2} into the following form,
\begin{align}
    \Delta\ep^1(R) &= \frac{\gamma}{R^3+R_0^3}, \label{Dep-3}
\end{align}
where we introduce another parameter $R_0$ such that the $R=0$ limit is correct.

\subsection{Further analysis on our model}

In the literature about Hubbard model, the ratio between Hubbard $U$ and the hopping integral $t$ is a critical factor that determines how strongly correlated a system is. This, however, is not directly applicable in our case since our Hubbard model is asymmetric, with different Hubbard interactions as well as different on-site energies. Here we perform the following analysis on our reduced Hamiltonian,
\begin{equation}
\bm {\bar H}_e=
  \begin{bmatrix}
    U_1 + \Delta \ep(R) & -\sqrt 2t(R) & 0  \\
    -\sqrt 2t(R) & 0 & -\sqrt 2t(R) \\
    0 & -\sqrt 2t(R) & U_2 - \Delta \ep(R)
  \end{bmatrix}.
\end{equation}
Note that $\Delta \ep(R)>0$ and $U_1+\Delta\ep(R)$ is much greater than the rest of the terms for all $R$. Therefore in the electronic ground state, the doubly occupied state on site 1 (Li atom) is unlikely to occur. This has been verified by Figure 1 of the main text, where $|c_{\rm {Li}^{-}\cdots \rm {F}^{+}}|^2$ is negligible.
Thus the $3\times 3$ matrix can reduce to the following $2\times 2$ matrix effectively.
\begin{equation}
\bm {\bar H}_{\rm eff}=
  \begin{bmatrix}
    0& -\sqrt 2t(R)  \\
    -\sqrt 2t(R) & U_2 - \Delta \ep(R)
  \end{bmatrix},
\end{equation}
where electron hopping takes place between the neutral and the ionic states, with the effective hopping integral being $\sqrt 2 t(R)$. Here the neutral state has zero Hubbard interaction, whereas the ionic state has an effective Hubbard interaction of $\tilde U_2(R) = U_2 - \Delta \ep(R)$.
Moreover, $\tilde U_2(R)$ experiences a sign change from negative to positive as $R$ increases. The critical $R_c$ where $\tilde U_2$ reaches zero corresponds to the charge transfer position under BO approximation. Therefore, the ratio of $q(R)=\frac{\tilde U_2(R)}{\sqrt 2 t(R)}$ is the critical factor that sensors the magnitude of correlation.

To see how strongly correlated our system is, we compute the natural occupation numbers as functions of $R$. If the system is non-interacting, the electronic wavefunction comes from a determinant and thus the natural occupation numbers are 1 and 0. If the system is strongly correlated, we should obtain fractional natural occupation numbers. In our case, let our electronic wavefunction be expanded in terms of basis states as
\begin{equation}
    \Phi = c_1 \varphi_1 + c_2 \varphi_2 + c_3 \varphi_3,
\end{equation}
where $\varphi_1 = |1_\ua 1_\da\rangle$, $\varphi_2 = \frac{1}{\sqrt 2}(|1_\ua 2_\da\rangle-|1_\da 2_\ua\rangle$), and $\varphi_3 = |2_\ua 2_\da\rangle$.
Then the electronic 1-body reduced density matrix (1-RDM) is given by
\begin{equation}
    \rho_{ij} = \langle \Phi | c^\dagger_{i\ua}c_{j\ua} | \Phi \rangle.
\end{equation}
Here we focus on one of the spin channels. In particular, $\rho_{ii}$ counts the total electron occupation on site $i$ and reads
\begin{align}
    \rho_{11} &= c_1^2+\frac{1}{2}c_2^2, \\
    \rho_{22} &= c_3^2+\frac{1}{2}c_2^2.
\end{align}
Moreover, by some calculation we obtain
\begin{align}
    \rho_{12} = \frac{1}{\sqrt 2} c_2(c_1+c_3).
\end{align}
Therefore we can compute the natural occupation numbers as given by the eigenvalues of the 1-RDM as
\begin{align}
    \lambda_{1,2} &= \frac{1}{2}\Big[(\rho_{11}+\rho_{22}\pm \sqrt{(\rho_{11}+\rho_{22})^2-4(\rho_{11}\rho_{22}-\rho_{12}^2)}\;\Big] \nonumber \\
    &=\frac{1}{2}\Big[1 \pm \sqrt {1-(c_2^2 - 2c_1c_3)^2} \;\Big].
\end{align}

\begin{figure}[t]
\includegraphics[width=\columnwidth]{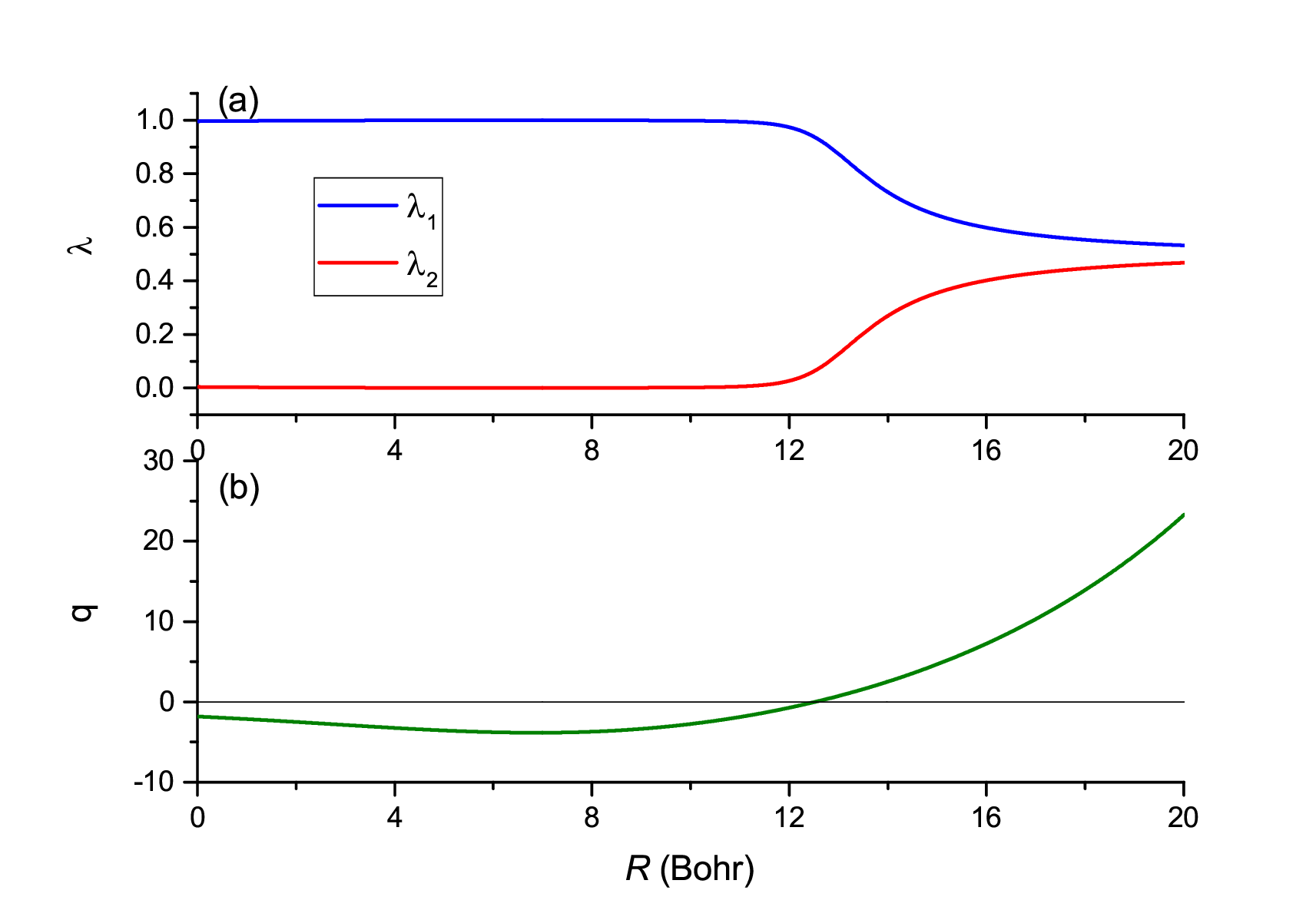}
\caption{
(a)Natural occupation numbers as functions of $R$; (b) the ratio between the effective Hubbard interaction and the hopping integral, $q(R)$.  \label{lambda}}
\end{figure}

In Fig \ref{lambda}(a) we show $\lambda_{1,2}$ as functions of $R$. As can be seen, for $R<R_c$, the $\lambda$'s are essentially 1 and 0, indicating that the system is weakly correlated. As $R$ increases from $R_c$, $\lambda_1$ and $\lambda_2$ start to approach each other and will finally reach identically 0.5 at the infinite limit. This suggests the amount of correlation starts to grow from the charge transfer position until becoming strongly correlated. This feature has also been manifested from the plot of $q(R)$, as shown in Fig \ref{lambda}(b). For $R<R_c$, $q(R)$ is negative due to the negative $\tilde U_2$; when $R>R_c$, $q(R)$ turns positive and keeps increasing, suggesting increasing amount of correlation, which is consistent with the conclusion established in the literature.

\subsection{B-spline basis set}
The B-spline basis functions used in our calculation have translational symmetries. Each spline function is defined by a piecewise function centered at $x_j$ with finite support $4h$. \cite{Hofmann01245321}
\begin{equation}
B_j(x)=
\left\{
\begin{array}{cl}
\frac{1}{4}(2+\frac{x-x_j}{h})^3, &\quad {\rm if}~-2<\frac{x-x_j}{h}\leqslant -1,\\
1-\frac{3}{2}(2+\frac{x-x_j}{h})^2-\frac{3}{4}(2+\frac{x-x_j}{h})^3,
 &\quad {\rm if}~-1<\frac{x-x_j}{h}\leqslant 0,\\
 1-\frac{3}{2}(2+\frac{x-x_j}{h})^2+\frac{3}{4}(2+\frac{x-x_j}{h})^3,
 &\quad {\rm if}\quad 0<\frac{x-x_j}{h}\leqslant 1,\\
 \frac{1}{4}(2-\frac{x-x_j}{h})^3, &\quad {\rm if}\quad 1<\frac{x-x_j}{h}\leqslant 2,\\
 0, &\quad {\rm otherwise}.
\end{array}
\right. \label{splie}
\end{equation}
In our calculation, we use 1600 such basis sets centered uniformly along the one dimensional real space from $x_0=0.2$ Bohr to $x_m=20.2$ Bohr, so that $h=0.0125$ Bohr.

\section{Supplemental derivations and results on density functionals}

\subsection{Some detailed derivation of the exact functional under BO approximation}

\begin{figure}[t]
\includegraphics[width=\columnwidth]{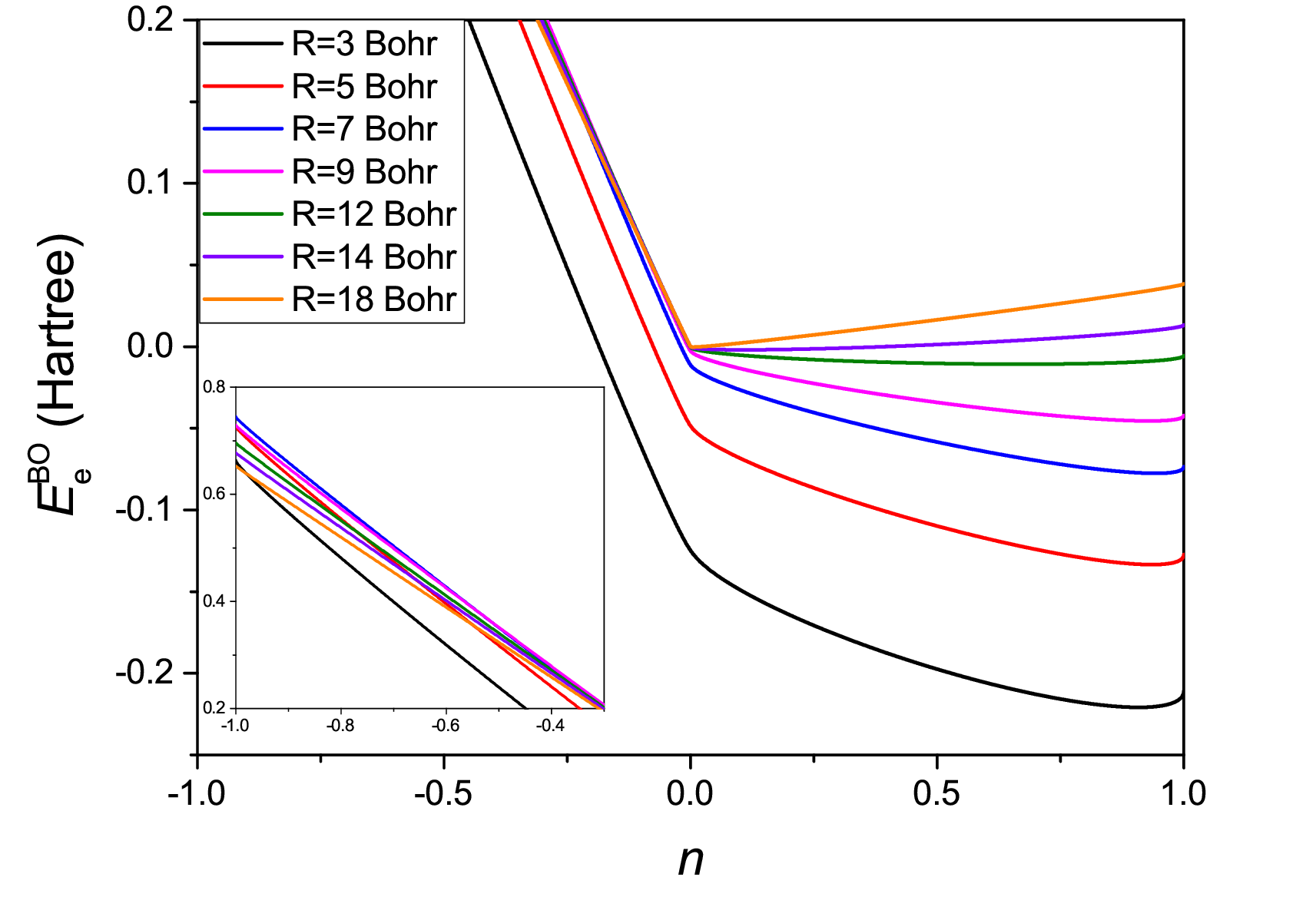}
\caption{
Exact functional under BO approximation.  \label{BO_func}}
\end{figure}

The exact functional under BO approximation is formally given by the Levy-constrained search formula,
\begin{align}
    E_e^{\rm BO}[n] &= \min_{\theta_1, \theta_2 \rightarrow n} E_e[\theta_1, \theta_2] \nonumber \\
   &= \min_{\theta_1, \theta_2 \rightarrow n} \Big\{-\sqrt 2 t \sin 2\theta_1 (\sin \theta_2 + \cos \theta_2)+ \frac{1}{2}\cos^2 \theta_1(\tilde U_1+\tilde U_2)\Big\} \nonumber \\
   &\quad + \frac{1}{2}n(\tilde U_2-\tilde U_1) + \tilde \epsilon_0. \label{Levy}
\end{align}
To accurately compute the functional for each $n$, we divide into two cases and simplify the expression as follows.
\newline
(1) $n=0$. Then either $\cos^2\theta_1=0$ or $\cos2\theta_2=0$.
\newline
(i) $\cos\theta_1=0$. Then $\sin2\theta_1=2\sin \theta_1\cos\theta_1=0$. the functional reduces to $E_e^{\rm BO}[n] = \tilde \epsilon_0$.
\newline
(ii) $\cos2\theta_2=0$. Then $2\theta_2=m\pi+\frac{1}{2}\pi \; \Longrightarrow \theta_2 = \frac{1}{2}m\pi + \frac{\pi}{4}$. Thus $\sin\theta_2+\cos\theta_2 = \sqrt 2 \sin(\theta_2+\frac{\pi}{4})=\sqrt 2 \sin(\frac{1}{2}m\pi + \frac{\pi}{2})$, which can be $0$ or $\pm \sqrt 2$. Then it is easy to see that
\begin{align}
    E_e^{\rm BO}[n] &= \min_{\theta_1, \theta_2 \rightarrow n} E_e[\theta_1, \theta_2]\nonumber \\
    &= \min_{\theta_1} \Big\{ -2 t |\sin 2\theta_1| + \frac{1}{2}\cos^2 \theta_1(\tilde U_1+\tilde U_2) \Big\}+\tilde \epsilon_0 \nonumber \\
    &= \min_{\theta_1} \Big\{-2 t |\sin 2\theta_1| + \frac{1}{4}\cos2 \theta_1(\tilde U_1+\tilde U_2)+ \frac{1}{4}(\tilde U_1+\tilde U_2)\Big\}+\tilde \epsilon_0 \nonumber \\
    &= -\sqrt{4t^2+\frac{1}{16}(\tilde U_1+\tilde U_2)^2}+ \frac{1}{4}(\tilde U_1+\tilde U_2)+ \tilde \epsilon_0. \label{energy1}
\end{align}
The minimizer is achieved when $\cos 2\theta_1 = -\frac{\frac{1}{4}(\tilde U_1+\tilde U_2)}{\sqrt{4t^2+\frac{1}{16}(\tilde U_1+\tilde U_2)^2}}$.
It is easy to see that $E_e^{\rm BO}[0]<E_e^{\rm BO, approx}[0]=\tilde \ep_0$.
\newline
(2) $n\neq 0$. Then the total electronic energy can be re-expressed in terms of $n$ and $\theta_1$. In \Eq{Levy}, one can eliminate the $\theta_2$ dependence and substitute it by a function of $n$ and $\theta_1$. Note that the only term that depends on $\theta_2$ is
$s=\sin \theta_2 + \cos\theta_2$.
Since $\cos 2\theta_2 = \frac{n}{\cos^2\theta_1}$, we try to express $s$ in terms of $\cos 2\theta_2$.
\begin{equation}
    s^2 = 1+\sin2\theta_2, \qquad \Longrightarrow (s^2-1)^2 = \sin^2 2\theta_2 = 1-\cos^2 2\theta_2=1-\frac{n^2}{\cos^4\theta_1}.
\end{equation}
Thus
\begin{equation}
    s = \pm \sqrt{1\pm \sqrt{1-\frac{n^2}{\cos^4\theta_1}}}.
\end{equation}
Denote the four roots of $s$ as $s_1, s_2, s_3$ and $s_4$, with the sign of $(++), (+-),(-+)$ and $(--)$, respectively. Then the ground state functional reduces to
\begin{align}
    E_e^{\rm BO}[n]  &= \min_{\theta_1} \Big\{\sqrt 2 t E_1[n,\theta_1]+ \frac{1}{2}\cos^2 \theta_1(\tilde U_1+\tilde U_2)\Big\}+ \frac{1}{2}n(\tilde U_2-\tilde U_1) + \tilde \epsilon_0, \label{Energy2}
\end{align}
where
\begin{equation}
    E_1[n,\theta_1] = \min\{-s_1 \sin2\theta_1, -s_2 \sin2\theta_1, -s_3 \sin2\theta_1, -s_4 \sin2\theta_1\}.
\end{equation}
Note that $s_3=-s_1$, and $s_4=-s_2$. Using this property, one can simplify the above expression of $E_1$ as
\begin{equation}
    E_1[n,\theta_1] = \min\{-s_1 |\sin2\theta_1|, -s_2 |\sin2\theta_1| \} = -s_1|\sin2\theta_1|.
\end{equation}
Thus \Eq{Energy2} translates to
\begin{align}
    E_e^{\rm BO}[n]  &= \min_{\theta_1} \Big\{-\sqrt 2 t|\sin2\theta_1|s_1 + \frac{1}{2}\cos^2 \theta_1(\tilde U_1+\tilde U_2)\Big\}+ \frac{1}{2}n(\tilde U_2-\tilde U_1) + \tilde \epsilon_0 \nonumber \\
    &= \min_{\theta_1} \Big\{-2\sqrt 2 t \sqrt{\frac{1}{4}(1-\cos^2 2\theta_1)\Big[1+ \sqrt{1-\frac{n^2}{\cos^4\theta_1}}\;\Big]}+ \frac{1}{2}\cos^2 \theta_1(\tilde U_1+\tilde U_2)\Big\} \nonumber \\
    &\quad + \frac{1}{2}n(\tilde U_2-\tilde U_1) + \tilde \epsilon_0. \label{Energy3}
\end{align}
Now define a variable $u$ that measures the deviation of the density from the boundary of its allowed domain ($|n|\leqslant \cos^2\theta_1$) as
\begin{equation}
    u = \sqrt{1-\frac{|n|}{\cos^2\theta_1}}\;.
\end{equation}
Then $0\leqslant u \leqslant 1$, and
\begin{equation}
    \cos^2\theta_1 = \frac{|n|}{1-u^2}. \label{costheta1sq}
\end{equation}
Moreover,
\begin{equation}
    \cos2\theta_1 = 2\cos^2\theta_1-1 = \frac{2|n|}{1-u^2}-1. \label{cos2theta1}
\end{equation}
Substituting \Eqs{costheta1sq}--\eqref{cos2theta1} into \Eq{Energy3}, we have
\begin{align}
    E_e^{\rm BO}[n] &= \min_{0\leqslant u \leqslant \sqrt{1-|n|}} \Big\{-2\sqrt 2 t\sqrt{(1-\frac{|n|}{1-u^2})\frac{|n|}{1-u^2}\Big[1+ u\sqrt{2-u^2}\;\Big]}\nonumber \\
    &\quad + \frac{|n|}{2(1-u^2)}(\tilde U_1+\tilde U_2) \Big\}+ \frac{1}{2}n(\tilde U_2-\tilde U_1) + \tilde \epsilon_0. \label{exact_func}
\end{align}
Given $n$, with an initial guess of $u_0=0$, one can perform line search to accurately obtain the minimizer. In Figure 2 of the main text, we have shown the plots of $E_e^{\rm BO}$ restricted to $n\in[0,1]$.
Here in Fig~\ref{BO_func}, we supplement with the plot for the whole range of $n\in[-1,1]$.

\begin{figure}[H]
\includegraphics[width=\columnwidth]{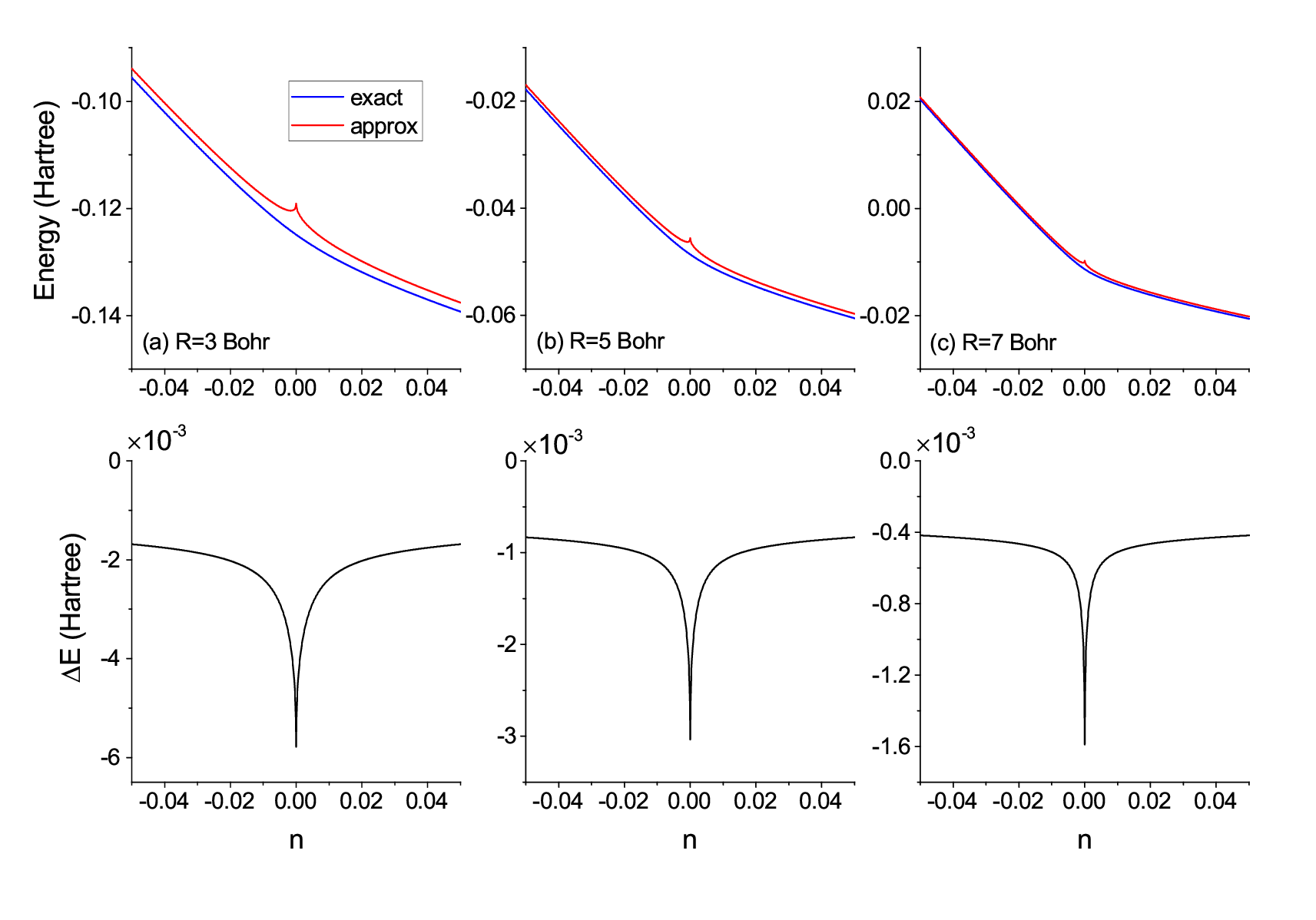}
\caption{
Comparison between the exact BO functional $E_e^{\rm BO}[n]$, the approximate BO functional $E_e^{\rm BO, approx}[n]$ and their difference $\Delta E [n]= E_e^{\rm BO}[n] - E_e^{\rm BO, approx}[n]$.  \label{comp_func_BO}}
\end{figure}

In Fig~\ref{comp_func_BO}, we compare the approximate BO functional $E_e^{\rm BO, approx}[n]$ with the exact $E_e^{\rm BO}[n]$ for selected $R$. As can be seen, their difference is small and the $E_e^{\rm BO, approx}[n]$ is a very good approximation to $E_e^{\rm BO}[n]$.

\subsection{Quantum geometric scalar $g$ as a function of the density $n$ and auxiliary variable $u$}

The quantum geometric scalar $g$ is defined as
\begin{equation}
    g(R) = |\nabla \Phi|^2 = \sum_{k=1}^3 \Big(\frac{d c_k}{dR}\Big)^2,
\end{equation}
where $c_k$ is the $k$th component of the electronic wave function in its basis functions. In terms of $\theta_1$ and $\theta_2$, $g(R)$ reads
\begin{align}
    g(R) &= \Big[\frac{d}{dR}(\cos\theta_1\sin\theta_2)\Big]^2 + \Big[\frac{d}{dR}(\sin\theta_1)\Big]^2 + \Big[\frac{d}{dR}(\cos\theta_1\cos\theta_2)\Big]^2 \nonumber \\
    &= \Big(\frac{d\theta_1}{dR}\Big)^2 + \cos^2\theta_1 \Big(\frac{d\theta_2}{dR}\Big)^2. \label{g}
\end{align}
Now $\theta_1$ and $\theta_2$ are linked to $n$ and $u$ through the following transformation,
\begin{equation}
    \cos^2\theta_1 = \frac{n}{1-u^2}, \label{theta1}
\end{equation}
and
\begin{equation}
    \cos2\theta_2 = 1-u^2. \label{theta2}
\end{equation}
From \Eq{theta1}, taking the derivative with respect to $R$ on both sides, we have
\begin{equation}
    -2\sin\theta_1\cos\theta_1 \frac{d\theta_1}{dR} = \frac{d}{dR}\Big(\frac{n}{1-u^2}\Big).
\end{equation}
Thus,
\begin{align}
    \Big(\frac{d\theta_1}{dR}\Big)^2 &= \frac{1}{4\sin^2\theta_1\cos^2\theta_1}\Big[\frac{d}{dR}\Big(\frac{n}{1-u^2}\Big)\Big]^2 = \frac{1}{4(1-\cos^2\theta_1)\cos^2\theta_1}\Big[\frac{d}{dR}\Big(\frac{n}{1-u^2}\Big)\Big]^2 \nonumber \\
    &= \frac{1}{4(1-\frac{n}{1-u^2})\frac{n}{1-u^2}}\Big[\frac{d}{dR}\Big(\frac{n}{1-u^2}\Big)\Big]^2 = \frac{(1-u^2)^2}{4n(1-u^2-n)}\Big[\frac{d}{dR}\Big(\frac{n}{1-u^2}\Big)\Big]^2. \label{dtheta1}
\end{align}
Similarly, taking the derivative with respect to $R$ on both sides of \Eq{theta2} leads to
\begin{equation}
    -2\sin2\theta_2\frac{d\theta_2}{dR} = -2u\frac{du}{dR}.
\end{equation}
It follows that
\begin{align}
    \Big(\frac{d\theta_2}{dR}\Big)^2 &= \frac{u^2}{\sin^2 2\theta_2}\Big(\frac{du}{dR}\Big)^2= \frac{u^2}{1-\cos^2 2\theta_2}\Big(\frac{du}{dR}\Big)^2 \nonumber \\
    &= \frac{u^2}{1-(1-u^2)^2}\Big(\frac{du}{dR}\Big)^2 = \frac{1}{2-u^2}\Big(\frac{du}{dR}\Big)^2. \label{dtheta2}
\end{align}
Now substituting \Eq{theta1}, \eqref{dtheta1} and \eqref{dtheta2} into \Eq{g}, we obtain
\begin{align}
    g(R) &= \frac{(1-u^2)^2}{4n(1-u^2-n)}\Big[\frac{d}{dR}\Big(\frac{n}{1-u^2}\Big)\Big]^2 + \frac{n}{(1-u^2)(2-u^2)}\Big(\frac{du}{dR}\Big)^2 \nonumber \\
    &= C_{nn}\Big(\frac{dn}{dR}\Big)^2 + C_{uu} \Big(\frac{du}{dR}\Big)^2 + C_{nu}\frac{dn}{dR}\cdot\frac{du}{dR},
    \label{g-1}
\end{align}
where by simple algebra one can arrive at
\begin{equation}
    C_{nn} = \frac{1}{4n(1-u^2-n)},
\end{equation}
\begin{equation}
    C_{uu} = \frac{n(1+nu^2-n)}{(1-u^2)^2(1-u^2-n)(2-u^2)},
\end{equation}
and
\begin{equation}
    C_{nu} = \frac{u}{(1-u^2)(1-u^2-n)}.
\end{equation}

\begin{figure}[H]
\includegraphics[width=0.97\columnwidth]{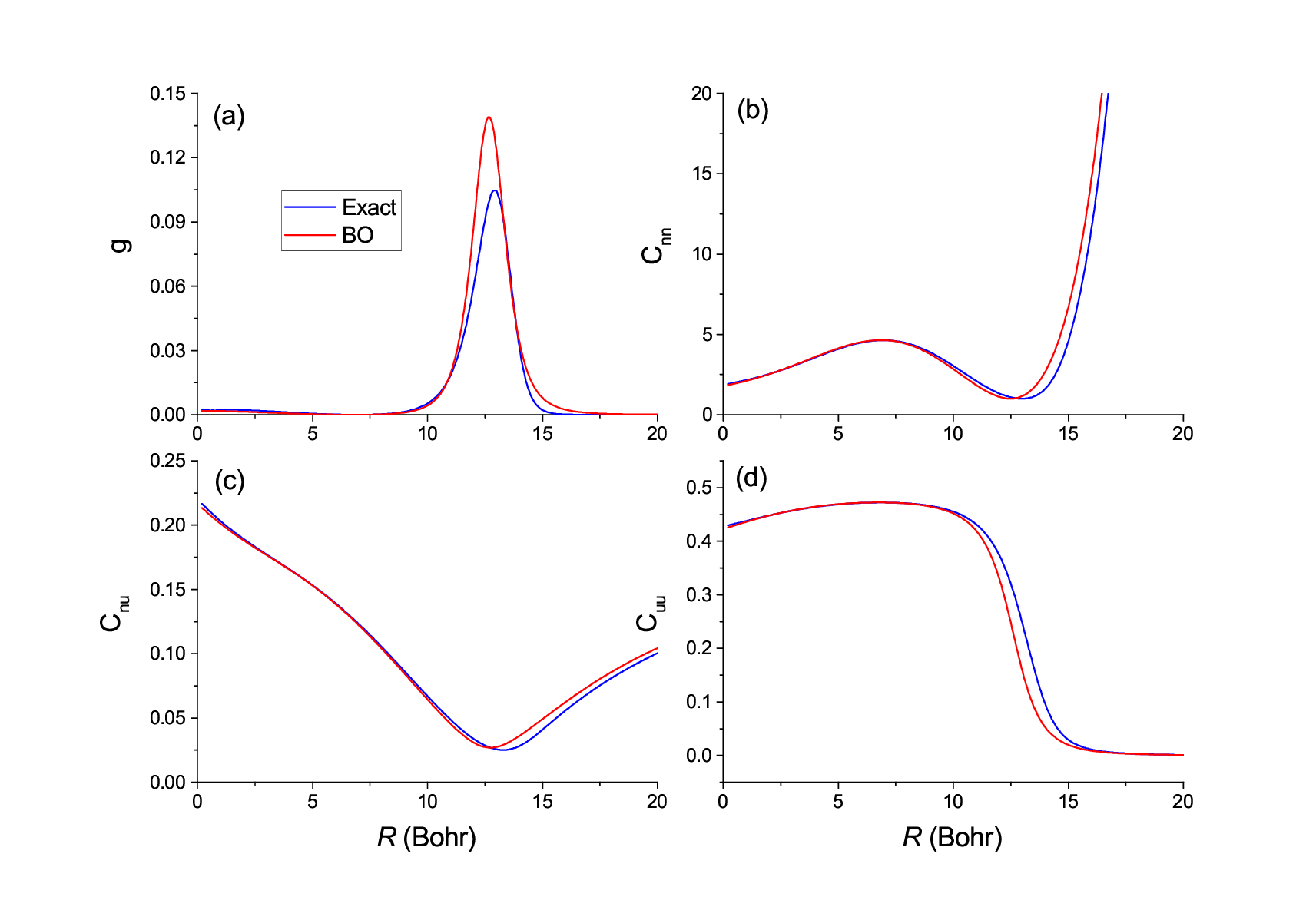}
\includegraphics[width=0.97\columnwidth]{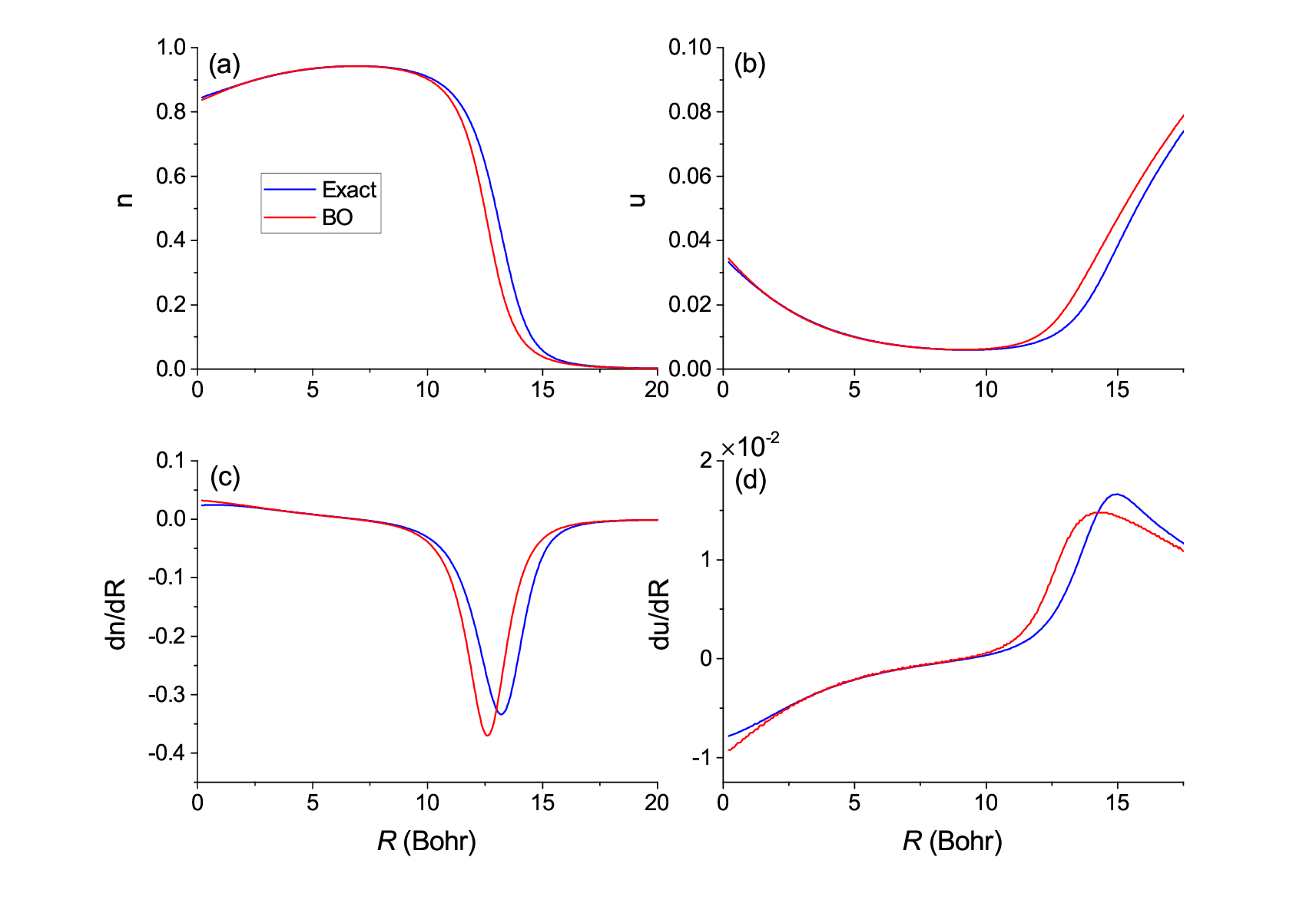}
\caption{
Upper panel: $g$, $C_{nn}$, $C_{nu}$ and $C_{uu}$ as functions of $R$. Lower panel: $n$, $u$, $\frac{dn}{dR}$ and $\frac{du}{dR}$ as functions of $R$. Comparison is made between the exact solution and the BO approximation. \label{Cnn}}
\end{figure}

\subsection{Computational details of local conditional density approximation}

For our 2-site Hubbard model, the Euler-Lagrange equation reads
\begin{align}
     \frac{\partial E_e^{\rm BO, approx}}{\partial n}& -\frac{\zeta}{M}\Big[\frac{1}{2}f'(n)\Big(\frac{dn}{dR}\Big)^2+f(n)\Big(\frac{d^2n}{dR^2}\Big)\Big] - \frac{1}{M}\frac{d\Big(\rm ln |\chi(R)|^2\Big)}{dR}f(n)\Big(\frac{dn}{dR}\Big)=0, \label{EL-n2}
\end{align}
where $\zeta=1$, and $f(n)=\frac{1}{4n(1-n)}$. Note that in the absence of the $\frac{1}{M}$-dependent terms, the remaining algebraic equation determines the ground state of the approximate BO functional, in particular,
\begin{align}
    -\sqrt 2 t(R) \frac{1-2n}{\sqrt{n(1-n)}}+\tilde U_2(R)=0. \label{algebraeq}
\end{align}
Here we have restricted to the $n>0$ branch because the global minimum of the BO (either exact or approximate) functional occurs in this branch.
The solution of \Eq{algebraeq} is given by
\begin{equation}
    n_0(R) = \frac{1}{2}\Big(1-\frac{q(R)}{\sqrt{q(R)^2+4}}\;\Big),
\end{equation}
where $q(R) = \frac{\tilde U_2(R)}{\sqrt 2 t(R)}$.

Now denote the $\frac{1}{M}$-dependent term in \Eq{EL-n2} as $A[n(R)]$. With that term present, we solve the equation iteratively as follows. First, we use $n_0(R)$ as an initial guess and compute its $R$-space derivatives, and obtain $A_0(R)=A[n_0(R)]$. Treating $A_0(R)$ as a correction to $\tilde U_2(R)$, we then update $q(R)$ by $q_1(R)=\frac{\tilde U_2(R)+A_0(R)}{\sqrt 2 t(R)}$. Substituting $q(R)$ by $q_1(R)$, we obtain the solution to the corrected algebraic equation \eqref{algebraeq}, denote it by $\tilde n_1(R)$. Next we generate the density for the next iteration through $n_1(R)=(1-\xi)n_0(R)+ \xi \tilde n_1(R)$, where $\xi$ is a damping factor introduced to guarantee convergence. We iterate this process until $\max_{R} \frac{1}{\xi}|n_k(R)-n_{k-1}(R)|$ drops below a certain threshold, say $10^{-5}$, where we believe convergence is reached.

\begin{figure}[H]
\includegraphics[width=\columnwidth]{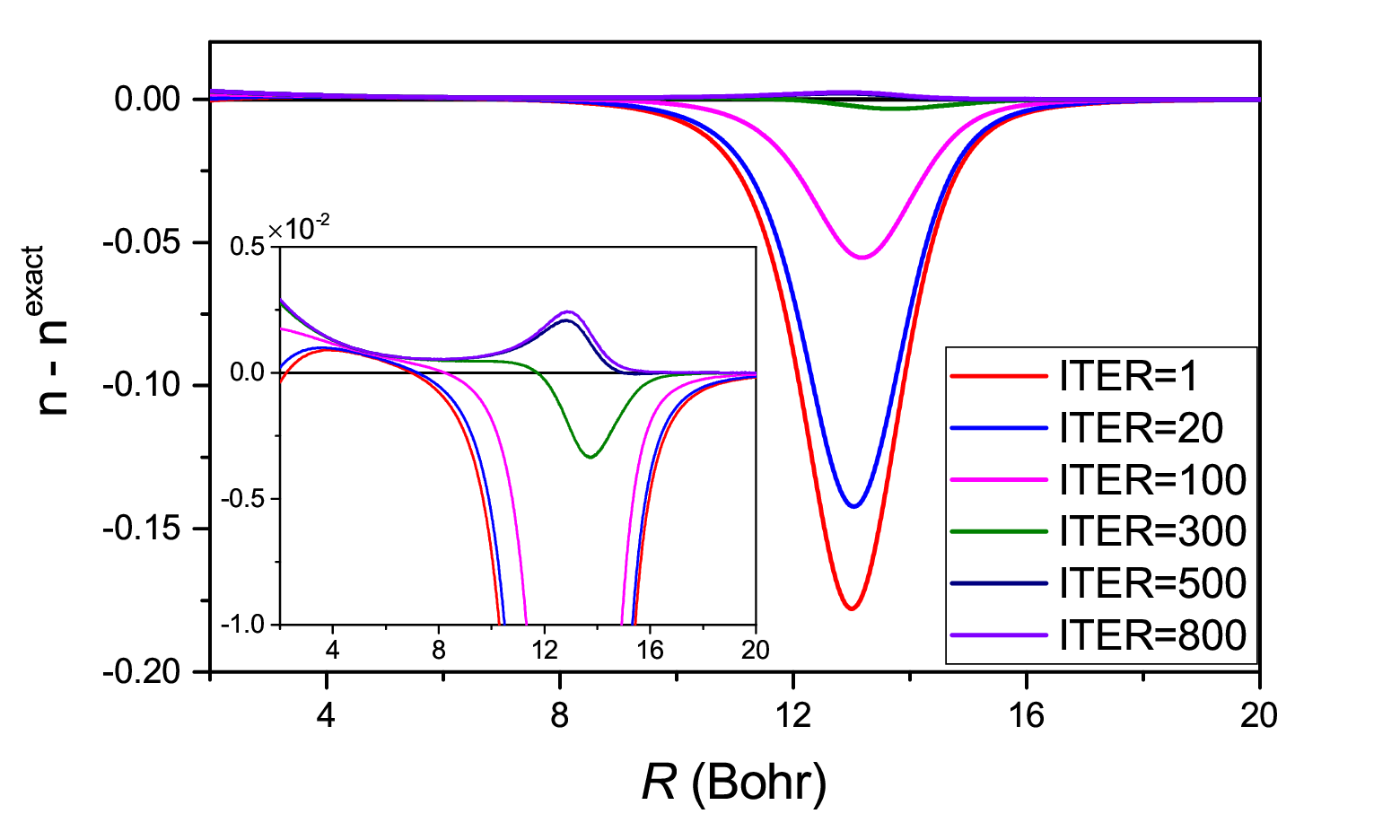}
\caption{
Convergence process of the conditional electronic density $n(R)$ in solving the Euler-Lagrange equation ($\zeta=1$).
Shown in the inset is the enlarged region near $n(R)=n^{\rm exact}(R)$. \label{LCDA_conv}}
\end{figure}

\begin{figure}[H]
\includegraphics[width=\columnwidth]{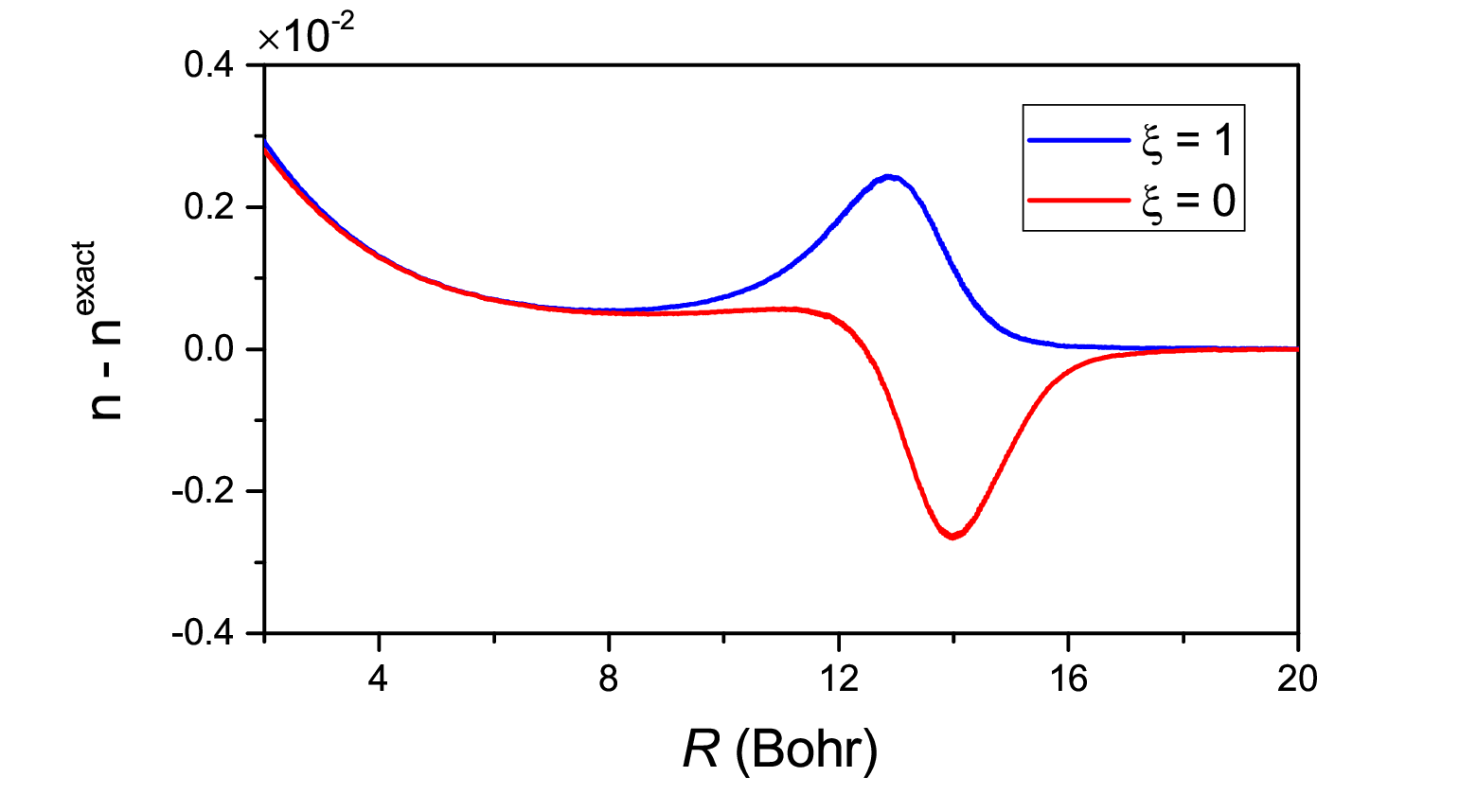}
\caption{
Comparison between solutions of the Euler-Lagrange equation relative to the exact density with $\zeta=1$ and $\zeta=0$. \label{comp_LCDA}}
\end{figure}

In Fig \ref{LCDA_conv} we present the convergence process when solving the Euler-Lagrange equation. As can be seen, although it takes hundreds of iterations, the $n(R)$ will finally converge to a density that is close to $n^{\rm exact}(R)$ with deviation on the magnitude of $10^{-3}$. As a remark, we note among the two terms in \Eq{EL-n2} that depend on $\frac{1}{M}$, the term involving the gradient of $\rm ln|\chi|^2$ has the dominant effect. This has been verified by solving the equation without the other term (i.e., setting $\xi=0$), where the solution is similar and also close to $n^{\rm exact}$, see Fig \ref{comp_LCDA}.

\begin{figure}[H]
\includegraphics[width=\columnwidth]{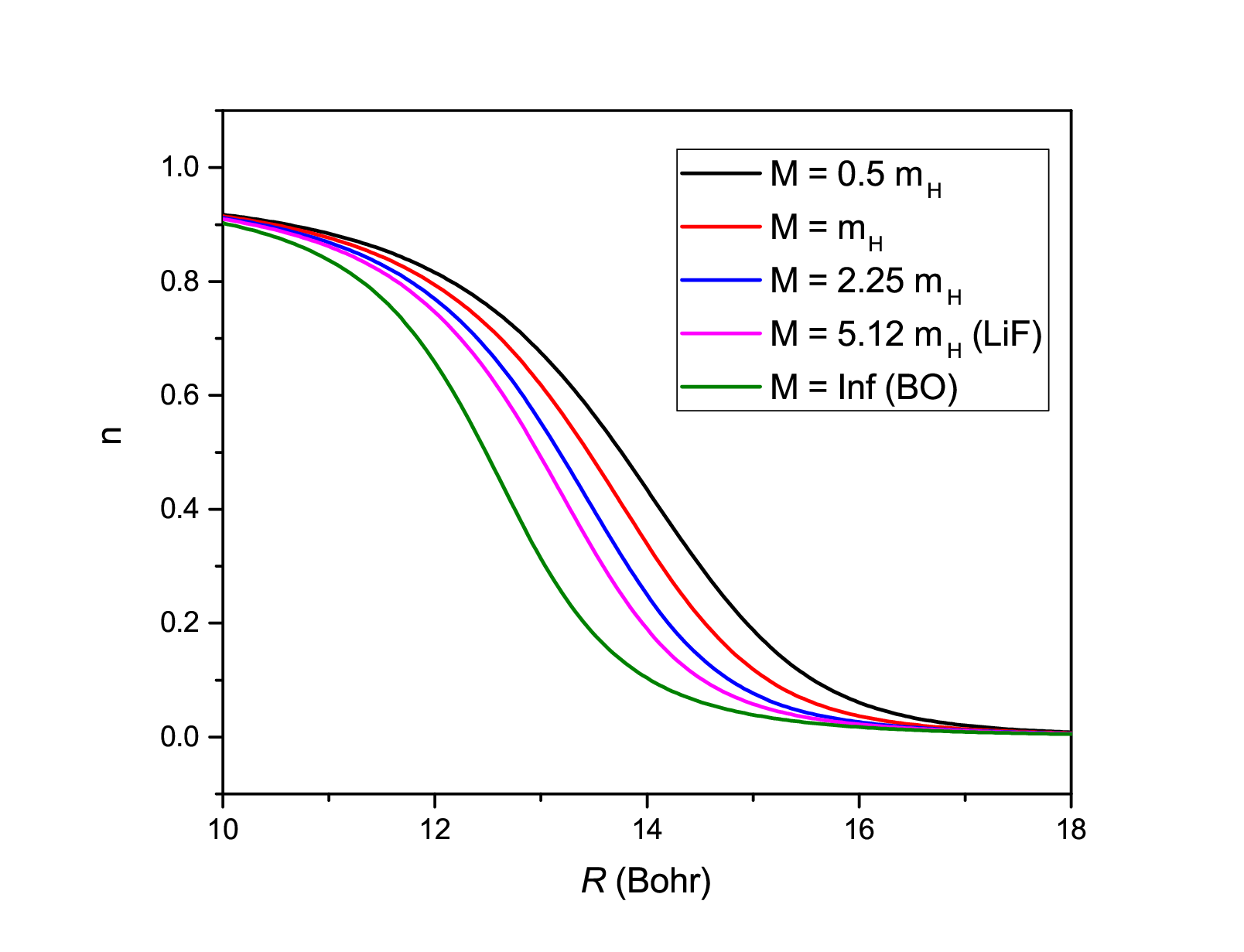}
\caption{
Comparison between the exact densities for our 2-site Hubbard model with different reduced mass. \label{comp_mass}}
\end{figure}

As an additional remark, we note that the non-adiabatic correction in \Eq{EL-n2} is proportional to $\frac{1}{M}$. Therefore, with smaller mass, we shall observe larger shift of the charge transfer position. In Fig~\ref{comp_mass}, we present the exact densities for our 2-site Hubbard model with different reduced masses in unit of the hydrogen mass. As can be seen, the trend is as expected. If $M$ equals the hydrogen mass, the right shift from the BO prediction can be as large as 1 Bohr. Moreover, the transition occurs in a milder way.

\section{Kohn-Sham equations for the 2-site Hubbard model}
\subsection{Kohn-Sham equation for the BO functional}
The Kohn-Sham equation for our model reduces to the following eigenvalue equation of a $2\times2$ matrix,
\begin{equation}
    \begin{bmatrix}
    -\frac{1}{2}\Delta v_s & -t  \\
    -t & \frac{1}{2}\Delta v_s
  \end{bmatrix}
  \begin{bmatrix}
    \sin\frac{\theta}{2} \\
    \cos\frac{\theta}{2}
  \end{bmatrix} = \epsilon \begin{bmatrix}
    \sin\frac{\theta}{2} \\
    \cos\frac{\theta}{2}
  \end{bmatrix}. \label{KS0}
\end{equation}
Here $\Delta v_s$ is the KS effective potential bias over the two sites.
%For the occupied state, $\epsilon = -\sqrt{t^2+\Delta v_s^2}$.
Without loss of generality, we assume $\sin\frac{\theta}{2}$ and $\cos\frac{\theta}{2}$ are all non-negative (bonding state). Since $n=\cos^2\frac{\theta}{2}-\sin^2\frac{\theta}{2}=\cos\theta$, we have $\cos\frac{\theta}{2} = \sqrt{\frac{1+n}{2}}$ and $\sin\frac{\theta}{2} = \sqrt{\frac{1-n}{2}}$.
%It follows that $\Delta v_s$ satisfies the following equation,
%\begin{equation}
%    \Delta v_s \sqrt{\frac{1-n}{2}} - t \sqrt{\frac{1+n}{2}} = -\sqrt{t^2+\Delta v_s^2}\sqrt{\frac{1-n}{2}}.
%\end{equation}
Therefore, we can solve for $\Delta v_s$ as
\begin{equation}
    \Delta v_s = -\frac{2nt}{\sqrt{1-n^2}}.
\end{equation}
This is the mapping between the density and the KS potential.

In the following, we explicitly write down the formula of each component of the exact KS energy functional.
The non-interacting kinetic energy is given by
\begin{equation}
    T_s = 2[\sin\frac{\theta}{2}, \cos\frac{\theta}{2}]
    \begin{bmatrix}
    0 & -t  \\
    -t & 0
  \end{bmatrix}
  \begin{bmatrix}
    \sin\frac{\theta}{2} \\
    \cos\frac{\theta}{2}
  \end{bmatrix} = -4t \sin\frac{\theta}{2}\cos\frac{\theta}{2} = -2t\sqrt{1-n^2}.
\end{equation}
Here the factor 2 is due to the double occupancy of the KS orbital. It is easy to verify that
\begin{equation}
    \frac{d T_s[n]}{dn} = -\Delta v_s.
\end{equation}
Thus solving for the ground state density through
\begin{equation}
    \frac{d E_e^{\rm BO}[n]}{dn} = 0
\end{equation}
is equivalent to solving the KS equation of \eqref{KS0} with
\begin{equation}
    \frac{d }{dn}\Big(E_e^{\rm BO}[n]-T_s[n]\Big) = \Delta v_s. \label{Delta-vs}
\end{equation}
Note here $E_e^{\rm BO}-T_s$ defines the sum of the electron-nuclear interaction energy $E_{en}$ and the Hartree exchange-correlation energy $E_{hxc}$. In particular,
\begin{equation}
    E_{en} = -n\Delta \epsilon = n(\epsilon_2 - \epsilon_1),
\end{equation}
and invoking the definition of $E_e^{\rm BO}$ as in \Eq{exact_func}, we can write $E_{hxc}$ explicitly as
\begin{align}
    E_{hxc} = E_e^{\rm BO} - E_{en} - T_s &= \min_{0\leqslant u \leqslant \sqrt{1-|n|}} \Big\{-2\sqrt 2 t\sqrt{(1-\frac{|n|}{1-u^2})\frac{|n|}{1-u^2}\Big[1+ u\sqrt{2-u^2}\;\Big]}\nonumber \\
    &\quad + \frac{|n|}{2(1-u^2)}( U_1+U_2) \Big\}+ \frac{1}{2}n( U_2- U_1) + \tilde \epsilon_0 - T_s - E_{en}. \label{exact_func1}
\end{align}
%As a remark, here $E^{BO} - E_{ext}$ defines the universal functional.
\subsection{Kohn-Sham equation for the exact factorization}

In the exact factorization, the total electronic part of the energy is approximated by
\begin{equation}
    E_e[n(R)] = \int |\chi(R)|^2 \Big[E_e^{\rm BO, approx}[n]+\frac{1}{2M}f(n)\Big(\frac{dn}{dR}\Big)^2\Big]dR,
\end{equation}
where the second term in the square bracket is the geometric contribution, expressed in terms of our local conditional density approximation (LCDA). Here $f(n)=\frac{1}{4n(1-n)}$.
%In practice, we can replace the exact BO functional $E^{\rm BO}$ by the approximate one $\bar E^{\rm BO}$.

Suppose $\chi(R)$ is given, the Euler-Lagrange equation leads to
\begin{align}
    \frac{\partial E_e^{\rm BO, approx}}{\partial n}-\frac{1}{M}\Big[\frac{1}{2}f'(n)\Big(\frac{dn}{dR}\Big)^2+f(n)\Big(\frac{d^2n}{dR^2}\Big)\Big]- \frac{1}{M}\frac{d\Big(\rm ln |\chi(R)|^2\Big)}{dR}f(n)\Big(\frac{dn}{dR}\Big)=0. \label{EL1}
\end{align}
Let
\begin{equation}
    v_{\rm geo}(R) = -\frac{1}{M}\Big\{\frac{1}{2}f'(n)\Big(\frac{dn}{dR}\Big)^2+f(n)\Big(\frac{d^2n}{dR^2}\Big) + \frac{d\Big(\rm ln |\chi(R)|^2\Big)}{dR}f(n)\Big(\frac{dn}{dR}\Big)\Big\},
\end{equation}
then it is easy to see that \Eq{EL1} is equivalent to the coupled Kohn-Sham equation (which involves the derivative with respect to $R$) of
\begin{equation}
    \begin{bmatrix}
    -\frac{1}{2}[\Delta v_s(R) + v_{\rm geo}(R)] & -t(R)  \\
    -t(R) & \frac{1}{2}[\Delta v_s(R) + v_{\rm geo}(R)]
  \end{bmatrix}
  \begin{bmatrix}
    \sin\frac{\theta(R)}{2} \\
    \cos\frac{\theta(R)}{2}
  \end{bmatrix} = \epsilon(R) \begin{bmatrix}
    \sin\frac{\theta(R)}{2} \\
    \cos\frac{\theta(R)}{2}
  \end{bmatrix}. \label{KS1}
\end{equation}
Here $v_{\rm geo}$ is the geometric correction to the KS potential.
Therefore, through approximating $E_e^{\rm BO}$ by $E_e^{\rm BO, approx}$, the solution of the Euler-Lagrange equation as shown in the main text is the same as the Kohn-Sham solution of \Eq{KS1}. In other word, it is possible to modify the Kohn-Sham equation to capture the non-adiabatic effects.